\documentclass[11pt,a4paper]{article}
\usepackage{jheppub}
\usepackage{textcomp}

\newcommand{\lag}{\mathcal{L}}

\newcommand{\muEW}{\mu_{\rm EW}}
\def\msbar{$\overline{\hbox{MS}}$}

\usepackage{multirow}
\usepackage{caption}
\usepackage{subcaption}
\usepackage{placeins}

\def\msbar{$\overline{\hbox{MS}}$}

\title{Threshold corrections in SMEFT}
\preprint{IPPP/26/05, KA-TP-02-2026, P3H-26-004}

\author[a]{Anke Biek\"otter,}
\author[b]{Livia E.~G.~Maskos,}
\author[b]{Benjamin D.~Pecjak}
\affiliation[a]{Institute for Theoretical Physics, Karlsruhe Institute of Technology, 76131 Karlsruhe, Germany}
\affiliation[b]{Institute for Particle Physics Phenomenology, 
Durham University,  Durham DH1 3LE, UK}

\abstract{
Threshold corrections provide a universal link between experimentally measured broken-phase parameters and
\msbar~renormalised symmetric-phase parameters used in SMEFT renormalisation-group (RG) analyses and matching to 
new-physics models. In this work we compute in analytic form the complete set of one-loop threshold corrections in dimension-six SMEFT, using two electroweak input schemes and including tadpole effects in the Fleischer--Jegerlehner (FJ) scheme.   As a by-product of the analysis, we obtain the full one-loop running of the strong coupling in SMEFT, including electroweak corrections, and the associated decoupling constants when the top quark and heavy electroweak bosons are integrated out.  Although generally moderate, the loop-level dimension-six corrections can be enhanced through tadpole effects and top-quark loops, and can shift symmetric phase parameters at the 5\% level under reasonable assumptions.  Our results constitute a process-independent ingredient which can be readily implemented in public RG-running and matching codes in SMEFT.

}

\begin{document}
\maketitle

\section{Introduction}
\label{sec:intro}
 
The Standard Model Effective Field Theory (SMEFT) \cite{Buchmuller:1985jz} is a widely used framework for analysing 
indirect effects of high-scale new physics on collider data. 
The general strategy is to express observables as an expansion in loops and operator 
dimensions within SMEFT, and then perform global fits to constrain the Wilson coefficients of 
higher-dimensional operators, see e.g.~\cite{Ellis:2018gqa, Dawson:2020oco, Bartocci:2024fmm, terHoeve:2025gey, deBlas:2025xhe}.

The characteristic energy of collider measurements used in such fits is typically
on the order of the electroweak (EW) scale, and the SMEFT
expansion of observables at that scale is performed using the broken-phase Lagrangian in the mass basis.
While the SMEFT Wilson coefficients and the strong-coupling constant $\alpha_s$ are renormalised in the \msbar~scheme, 
the remaining broken-phase parameters are typically renormalised on-shell in a particular EW input scheme,
thereby maintaining a direct link to experimental measurements.

Global fits enable model-independent analyses at the EW scale.
However, to extrapolate the bounds to energy scales $\Lambda_{\rm NP}$ characteristic of heavy new physics, 
one must perform renormalisation-group (RG) running within SMEFT, in which case the \msbar-renormalised
Lagrangian in its symmetric form is a convenient starting point. The gauge, Yukawa, and Higgs potential parameters
used in this symmetric-phase formulation are the same quantities that appear
in perturbative calculations of SMEFT Wilson coefficients in top-down approaches based on specific UV models, 
which have seen significant advances in recent years~\cite{deBlas:2017xtg,  Fuentes-Martin:2022jrf,Guedes:2023azv,Carmona:2021xtq,Guedes:2024vuf,Kraml:2025fpv}.
 
While the broken-phase parameters can be determined directly from 
experimental measurements at the EW scale, the symmetric-phase parameters cannot. 
Instead, by matching the Lagrangian across 
the unbroken and broken phases, one can derive relations between the 
two sets of parameters, allowing the symmetric-phase quantities to be expressed 
in terms of the broken-phase ones within a SMEFT expansion. The coefficients appearing 
in this expansion are commonly referred to as threshold corrections. These threshold 
corrections are obtained from a Lagrangian-level analysis and therefore constitute a 
universal, process-independent ingredient in RG analyses and matching computations in SMEFT.
 
The main goal of this paper is to calculate the threshold corrections to one loop in dimension-six SMEFT  within two
EW input schemes, namely those using  $\{M_W,M_Z,G_F\}$ or $\{M_W,M_Z,\alpha\}$ as inputs.
Our work builds on the current state of the art, which is two-loops within the SM  \cite{Kniehl:2015nwa} and tree-level in dimension-six 
SMEFT \cite{Alonso:2013hga}.  Although our computational procedure follows that used in SM analyses, new subtleties arise already at one loop.  For example, in both the SM and SMEFT, tadpole corrections contribute to the one-loop threshold corrections for the Higgs potential parameters. However, in SMEFT, they also affect the threshold corrections relating the SU(3) coupling $g_3$ in the unbroken phase to the strong-coupling constant $g_s$, unless the top quark and heavy EW bosons are decoupled. This leads us to perform the first calculation of one-loop decoupling constants for $g_s$ in SMEFT in the presence of EW corrections, extending the QCD results of
\cite{Duhr:2025yor} and providing a necessary element for consistently implementing experimental determinations 
of $\alpha_s(\mu)$ that assume SM-like running into SMEFT analyses.

The paper is organised as follows. Section~\ref{sec:preliminaries} defines the Lagrangian parameters in both the broken and unbroken phases, along with the expansion coefficients for threshold corrections in dimension-six SMEFT in the two electroweak input schemes. Section~\ref{sec:details} describes the computation of threshold corrections from counterterms, outlines cross-checks of our results, and discusses the renormalisation of the strong coupling constant in the broken phase, including the calculation of decoupling constants. A brief numerical analysis is presented in Section~\ref{sec:numerics}, followed by conclusions in Section~\ref{sec:conclusions}. Appendix~\ref{sec:threshold_corrections} provides analytic formulas for the threshold corrections.

\section{Preliminaries}
\label{sec:preliminaries}
 We write the dimension-six SMEFT Lagrangian as 
\begin{align}
\lag = \lag^{(4)} + \lag^{(6)}  ;  \quad \lag^{(6)} =  \sum_i C_i(\mu) \, Q_i(\mu) \, ,
\end{align} 
where $ \lag^{(4)}$ denotes the SM Lagrangian and $ \lag^{(6)}$ is the 
dimension-six Lagrangian with operators $Q_i$ given in the Warsaw basis~\cite{Grzadkowski:2010es}. The corresponding 
Wilson coefficients $C_i(\mu)$ are renormalised in the \msbar~scheme and are implicitly suppressed by two powers of the new  physics scale  $\Lambda$.  We shall often keep the $\mu$-dependence implicit, writing $C_i \equiv C_i(\mu)$.
The 59 independent dimension-six operators, which in general carry flavour indices, are listed and grouped into eight classes in Table~\ref{op59}.\footnote{We employ the symmetric basis for the Wilson coefficients. This means that for Wilson coefficients contributing to operators with two identical fermion bilinears, we define the Wilson coefficient of both flavour combinations and take into account their symmetry, e.g.\ $C_{\substack{ll\\1221}} + C_{\substack{ll\\2112}} = 2 \, C_{\substack{ll\\1221}} $.}
We define the covariant derivative as
\begin{align}
D_\mu  &= \left[ \partial_\mu + i g_3 T^a G_\mu^a + i \frac{g_2}{2}  \sigma^i W_\mu^i + i \mathbf{y} g_1 B_\mu \right]   \, ,
\end{align}
where the $\sigma_i$ are Pauli matrices. For the SM Higgs potential we follow the convention
of \cite{Jenkins:2013zja}, writing
 \begin{align}
 V = \lambda \left(H^\dagger H - \frac{1}{2}v^2 \right)^2 = \lambda (H^\dagger H)^2 
 - \mu_H^2 H^\dagger H  + {\rm constant} \, , 
 \end{align}
where $\mu_H^2 \equiv  \lambda v^2$.  Throughout this work, we truncate the SMEFT expansion of a given quantity to 
linear order in the dimension-six SMEFT Wilson coefficients.

In the formulation above, gauge invariance is manifest, and if $\mu_H^2$ were negative it would
correspond to a scenario with no spontaneous symmetry breaking -- in other words, the symmetric phase.
The set of Lagrangian parameters appearing in this formulation is particularly useful for studies of
RG evolution up to much higher scales and for matching with new-physics models. We shall refer to these
as ``symmetric-phase'' parameters, and write the full set as
\begin{align}
\label{eq:p_sym}
p_i^{\rm sym}(\mu) \equiv p_i^{\rm sym} \in \{ g_1,  g_2, \lambda, \mu_H^2,y_t,g_3 \} .
\end{align}
We have approximated all Yukawa couplings as zero apart from that of the top quark, $y_t$, and
made explicit that these symmetric-phase parameters are defined in the \msbar{} scheme
and therefore depend on the renormalisation scale $\mu$.

Consider now the broken-phase formulation, which makes use of the physical mass spectrum arising when $\mu_H^2>0$.
After rotation to the mass basis, a convenient set of broken-phase parameters is
\begin{align}
\label{eq:p_broken}
p_i^{\rm broken} \in \{ M_W, M_Z, v_\sigma,m_H, m_t, g_s(\mu) \}  .
\end{align}
Here we have approximated the CKM matrix by the unit matrix and treated all fermions as massless except for
the top quark with mass $m_t$. Furthermore, using the parameter $v_\sigma \in \{v_\mu, v_\alpha\}$ allows one
to choose between the two EW input schemes defined in Table~\ref{tab:schemeDef}; results can be converted to
other schemes using expressions from \cite{Biekotter:2023xle, Biekotter:2023vbh}. The definitions of these parameters
are
\begin{align}
v_\mu = \left( \sqrt{2}G_F \right)^{-\frac12} \, , \qquad 
v_\alpha =\frac{2 M_W s_w}{\sqrt{4\pi \alpha(M_Z)}}\, ,
\end{align}
where here and below the cosine ($c_w$) and sine ($s_w$) of the Weinberg angle are defined as
\begin{align}
c_w = \frac{M_W}{M_Z} = \sqrt{1- s_w^2} \, . 
\end{align}

A short comment is in order concerning the definition of the strong coupling constant $g_s(\mu)$ in Eq.~(\ref{eq:p_broken}).
First,  it is renormalised in the \msbar~scheme and therefore depends on a renormalisation scale $\mu$,  whereas
all other broken-phase quantities are renormalised on-shell and thus $\mu$ independent. Second, experimental
extractions of this quantity are typically quoted at $\mu = M_Z$, with evolution to and from other scales calculated from the 
QCD $\beta$ function.  In SMEFT, however, we shall see that both top-quark loops and electroweak corrections modify this running 
already at one loop, which prevents a direct use of experimental extractions that assume purely SM evolution.
To avoid this complication, we define $g_s(\mu)$ in a five-flavour version of QED$\times$QCD. In this set-up the running
is SM-like,  while the effects of heavy particles are taken into  account through decoupling constants.  
An analogous situation arises for the effective on-shell fine structure constant $\alpha(M_Z)$, 
whose one-loop decoupling constants in SMEFT were derived in \cite{Cullen:2019nnr}.

\begin{table}
	\centering
	\begin{tabular}{c | l}
	 scheme  & inputs   \\ \hline	
		 $v_\mu$ &  $G_F$, $M_W$, $M_Z$   \\
		 $v_\alpha$ &  $\alpha(M_Z)$, $M_W$, $M_Z$				 
	\end{tabular}
	\caption{\label{tab:schemeDef} Nomenclature for the EW input schemes considered in this work.}
\end{table} 

While the broken-phase parameters can be determined from, and subsequently
used to describe, experimental measurements at the EW scale, it is the symmetric-phase parameters which are used 
at much higher scales $\mu\sim \Lambda$, whether to connect with calculations of 
the Wilson coefficients $C_i$ at that scale in explicit models, or to address phenomenologically
interesting questions such as the high-scale stability of the EW vacuum \cite{Degrassi:2012ry, Buttazzo:2013uya,Bednyakov:2015sca}. In order to shift
between the two sets of Lagrangian parameters at a matching scale $\mu_{\rm EW}$ in a continuous
fashion, one must calculate the relations between them as a SMEFT expansion in loops and 
operator dimension.  We write these relations, up to one loop in dimension-six SMEFT, in the form
\begin{align}
\label{eq:threshold_expansion}
g_1 & = \frac{2M_Z s_w}{v_\sigma}\left[1 + v_\sigma^2  \delta g_1^{(6,0,\sigma)} +
\frac{1}{v_\sigma^2} \delta g_1^{(4,1,\sigma)} +\delta g_1^{(6,1,\sigma)}      \right] \, , \nonumber \\
g_2 &= \frac{2M_W}{v_\sigma}\left[1 + v_\sigma^2 \delta g_2^{(6,0,\sigma)} +
\frac{1}{v_\sigma^2}  \delta g_2^{(4,1,\sigma)} + \delta g_2^{(6,1,\sigma)}      \right] \, , \nonumber \\
g_3 &= g_s\left[1 +v_\sigma^2 \delta g_3^{(6,0)}  + \delta g_3^{(4,1)} +\delta g_3^{(6,1,\sigma)}  \right] \, , \nonumber \\
y_t & = \frac{\sqrt{2} m_t}{v_\sigma}\left[1 + v_\sigma^2  \delta y_t^{(6,0,\sigma)} +
\frac{1}{v_\sigma^2} \delta y_t^{(4,1,\sigma)} +\delta y_t^{(6,1,\sigma)}   + 
\delta y_t^{\alpha_s(4,1)}  +\delta y_t^{\alpha_s(6,1,\sigma)} \right] \, ,  \nonumber \\
\lambda & = \frac{m_H^2}{2v_\sigma^2}\left[1 + v_\sigma^2  \delta \lambda^{(6,0,\sigma)} +
\frac{1}{v_\sigma^2} \delta \lambda^{(4,1,\sigma)} +\delta \lambda^{(6,1,\sigma)}      \right] \, ,  \nonumber \\
\mu_H^2 & = \frac{m_H^2}{2}\left[1 + v_\sigma^2  \left[\delta\mu_H^2\right]^{(6,0,\sigma)} +
\frac{1}{v_\sigma^2} \left[\delta\mu_H^2\right]^{(4,1)} +\left[\delta\mu_H^2\right]^{(6,1,\sigma)}    \right] \, .
\end{align}
The superscripts on the expansion coefficients 
$\delta g_1^{(i,j,\sigma)}$ label the dimension-$i$ contribution at $j$-loop order in the EW input scheme using $v_\sigma$ as an input, and similarly for those on $\delta g_2$, etc. In cases
where the coefficient does not depend on the scheme we have dropped the superscript $\sigma$,
and in the case of $y_t$ we have separated out the EW and QCD contributions, the latter being
labeled with an additional superscript $\alpha_s$. 

The expansion coefficients in Eq.~(\ref{eq:threshold_expansion}), which are finite and 
gauge invariant, are referred to as  ``threshold corrections'' in the literature. They have been
calculated up to two loops in the pure SM in \cite{Kniehl:2015nwa}, and to calculate them at one loop in dimension-six  SMEFT is the main goal of the present work.  We outline some aspects of this calculation in the following section.

\section{Calculating the threshold corrections}
\label{sec:details}
In order to determine the threshold corrections in SMEFT,  we use as a starting point
the following tree-level relations between the parameters of the broken and symmetric phases, which can be derived from results in  \cite{Alonso:2013hga}: 
\begin{align}
\label{eq:tree_relations}
g_1 & = \frac{2 M_Z s_w}{v_T}\left[1- v_T^2\left(\frac{1}{4s_w^2} C_{HD} 
+ \frac{c_w}{s_w} C_{HWB} + C_{HB} \right) \right] \, , \nonumber  \\
g_2 & = \frac{2 M_W}{v_T}\left[1- v_T^2  C_{HW}  \right] \, , \nonumber \\
g_3 & = g_s \left[1 - v_T^2 C_{HG}\right] \, , \nonumber  \\
\lambda & = \frac{m_H^2}{2v_T^2}\left[1+v_T^2 \left( \frac{ 3v_T^2}{m_H^2} C_{H} 
+ \frac{1}{2}C_{HD} - 2 C_{H\Box}    \right) \right] \ ,\nonumber  \\
y_t & = \frac{\sqrt{2}m_t}{v_T} + \frac{1}{2} v_T^2 C_{\substack{uH\\33}} \ ,\nonumber  \\
\mu_H^2 & =  \frac{m_H^2}{2}\left[1+v_T^2 \left( \frac{3v_T^2}{2m_H^2} C_{H} 
+ \frac{1}{2}C_{HD} - 2 C_{H\Box}    \right) \right] \, .
\end{align}
The vacuum expectation value (vev) of the Higgs field in SMEFT is defined through 
$\langle H^\dagger H\rangle =v_T^2/2$.

Beyond tree level, the above relations hold for bare quantities. We can use them 
to calculate the threshold corrections by replacing such bare quantities by counterterms
plus their renormalised equivalents, and then solving for the renormalised
symmetric-phase parameters in terms of the broken-phase ones using a SMEFT expansion. 
The details of this procedure are worked out in Appendix~\ref{sec:threshold_corrections}.
The main ingredients are the SMEFT expansion coefficients of counterterms, which we turn to next.

For a generic renormalised parameter $X$, we define  
\begin{align}
\label{eq:param_CT}
X_0 = X\left(1+\Delta X\right) \, , 
\end{align}
where the subscript ``0'' refers to the bare quantity and $\Delta X$ are counterterms.  For 
the SMEFT Wilson coefficients, we use
\begin{align}
 C_{i,0} & = C_i + \Delta C_i , &  \Delta C_i & \equiv \frac{1}{2 \epsilon} \dot{C}_i \equiv \frac{1}{2 \epsilon} \frac{\text{d} C_i}{\text{d} \log \mu} \, ,
\end{align}
where $\epsilon = (4-d)/2$ is the dimensional regulator in 
$d$-dimensions.\footnote{To avoid notational clutter here and throughout the paper 
we omit the standard finite pieces which accompany divergent parts of counterterms in the \msbar~scheme and
cancel out of expressions for threshold corrections.} 
Furthermore, for broken-phase parameters renormalised on-shell, i.e. $X^{\rm O.S.}\in\{M_W,M_Z,v_\sigma,m_t,m_H\}$, we split the counterterms into divergent and finite
pieces in the dimensional regulator as
\begin{align}
\label{eq:finite_split}
\Delta X^{\rm O.S.}= \frac{\Delta X^{\rm O.S.}_{\rm div.}}{\epsilon} + \delta X^{\rm O.S.} \, .
\end{align}
Counterterms for the symmetric-phase parameters are all defined in the \msbar~scheme and contain
only divergent parts. After solving for the threshold corrections as functions of renormalised broken-phase
parameters, the divergent pieces of various counterterms cancel against each other,
and the final results are a function of the SMEFT expansion coefficients of the
$\delta X^{\rm O.S.}$.  

While this procedure for calculating the threshold corrections is fairly straightforward, 
its implementation relies on explicit results for the counterterms of the Wilson coefficients,
the symmetric-phase parameters, and the broken-theory parameters up to one loop in
dimension-six SMEFT.  Those for the Wilson coefficients have been obtained in \cite{Jenkins:2013zja,Jenkins:2013wua,Alonso:2013hga}\footnote{We use the electronic implementation in \texttt{DsixTools}~\cite{Celis:2017hod, Fuentes-Martin:2020zaz} as the 
$\dot{C}_i$ typically depend on a large number of Wilson coefficients.}, while those
for the symmetric-phase parameters were given in \cite{Jenkins:2013zja}.  
The counterterms for the full set of broken-theory parameters in SMEFT within the 
two different EW input schemes are not readily available in the literature, and
we have calculated them from scratch in the present work. This involves evaluating a 
number of Feynman  diagrams, each of which can receive contributions from multiple SMEFT operators. In order to perform these calculations, we have used a chain of automated tools.
It makes use of an in-house implementation of the SMEFT Lagrangian in \texttt{FeynRules}~\cite{Alloul:2013bka} in both unitary and Feynman gauge, a generation of Feynman 
diagrams with \texttt{Feynarts}~\cite{Hahn:1998yk} using the \texttt{Feynrules} output, 
and the calculation of loop diagrams in terms of Passarino-Veltmann~(PV) integrals using~\texttt{FormCalc}~\cite{Hahn:2016ebn,Hahn:1998yk,Hahn:2000kx}.  Our unitary-gauge results have 
been cross-checked using Feynman rules generated by \texttt{SMEFTsim}~\cite{Brivio:2017btx,Brivio:2020onw}, and various manipulations on PV integrals have been cross-checked between  \texttt{LoopTools}~\cite{Hahn:1998yk} and \texttt{PackageX}~\cite{Patel:2015tea}. 

We have performed a number of consistency checks on our results.  First, we have verified
the cancellation of all divergent contributions at the level of the threshold corrections, 
and the related fact that the $\mu$ dependence of the \msbar~renormalised symmetric phase parameters, calculated in terms of the broken-phase ones through threshold corrections, is
consistent with the renormalisation-group equations given in \cite{Jenkins:2013zja}.  
Second, we have performed all calculations in various gauges and checked that the threshold corrections are gauge independent. For the QCD corrections we utilised a general $R_\xi$ gauge,
while for electroweak corrections we have worked in both unitary and Feynman gauge. 
In the electroweak case tadpole contributions play an important role. We have treated tadpole counterterms in the FJ tadpole scheme~\cite{Fleischer:1980ub}, which amounts to including tadpole topologies in the Feynman diagrammatic calculations
of the counterterms \cite{Denner:2016etu},  and have confirmed the expected 
result that, after their inclusion, all counterterms are individually gauge invariant.  Finally,
the one-loop threshold corrections in the SM are a subset of our results, and we have
compared them with those given in \cite{Kniehl:2015nwa} in the $v_\mu$ scheme,
finding full agreement.

We end this section with a comment on the threshold corrections for the SU(3) 
coupling constant $g_3$. In the full electroweak SM, the symmetric-phase
parameter $g_3$ coincides with the corresponding broken-phase parameter at one loop. In the
SMEFT, however, a difference arises already at tree level, because the operator
$C_{HG}$ contributes to the broken-phase kinetic terms after electroweak symmetry
breaking.  In addition, the one-loop running of the strong coupling constant in SMEFT, which
we denote by $\overline{g}_s(\mu)$ to distinguish it from the five-flavour
QED$\times$QCD coupling $g_s(\mu)$ used in Eq.~(\ref{eq:p_broken}), receives
modifications from top-quark loops and weak corrections. We have computed this running
by renormalising the quark--gluon vertex, including tadpole contributions. 
Defining $\overline{\alpha}_s \equiv \overline{g}_s^2(\mu)/(4\pi)$, we obtain
\begin{align}
\label{eq:as_running}
\mu \frac{d}{d\mu}\overline{\alpha}_s = -2 \overline{\alpha}_s\left( \frac{\overline{\alpha}_s}{4\pi}\beta_0 -  \dot{g}_s^{(6,1)} \right) \, ,
\end{align}
where the one-loop SM result in our notation is $\beta_0=11 -2n_q/3$, with $n_q=6$ the number of
active quarks, and the dimension-six contribution is
\begin{align}
\label{eq:beta61}
16\pi^2 \dot{g}_s^{(6,1)} & = 
-4 \left( m_H^2+ \frac{6 M_W^4 + 3 M_Z^4 -12 m_t^4}{m_H^2}\right)C_{HG}
- 4\sqrt{2}\overline{g}_s m_t v_\sigma C_{\substack{uG\\33}}   \,.
\end{align}
The contribution proportional to $C_{\substack{uG\\33}}$ is a QCD effect arising from top-quark couplings to gluons
and agrees with the result obtained in \cite{Deutschmann:2017qum}, while that proportional to $C_{HG}$ is a pure weak  
effect and has not yet been considered in the literature. Higher-order QCD corrections have been considered in 
 \cite{Duhr:2025zqw, Duhr:2025yor}.
 
As mentioned in Section~\ref{sec:preliminaries},  dimension-six running effects can be avoided 
by using $g_s(\mu)$, whose RG equation takes the form of the SM piece of Eq.~(\ref{eq:as_running}) but with $n_q=5$.   
Up to corrections in light-quark masses,  this definition of the strong coupling is related to that in the full EW SMEFT 
by a decoupling constant 
$\zeta_{\alpha_s}$, which we expand to one-loop as
\begin{align}
\label{eq:alpha_decoup}
\overline{\alpha}_s =\left(1 +    \zeta_{\alpha_s}^{(4,1)} +  \zeta_{\alpha_s}^{(6,1)} \right)\alpha_s   \, .
\end{align}
The calculation of the decoupling constants follows the procedure and definitions outlined for the electric charge in
 \cite{Cullen:2019nnr}. It leads to the standard QCD result 
\begin{align}
\label{eq:zeta_as_41}
 \zeta_{\alpha_s}^{(4,1)} &= \frac{\alpha_s}{6\pi} \ln\frac{\mu^2}{m_t^2}  \, , 
 \end{align}
 while for the dimension-six piece we find
\begin{align}
\label{eq:zeta_as_61}
16\pi^2 \zeta_{\alpha_s}^{(6,1)} & = - 4\sqrt{2}g_s m_t v_\sigma \ln\left(\frac{\mu^2}{m_t^2}\right)
C_{\substack{uG\\33}} +\frac{4}{m_H^2} C_{HG}\bigg[12 m_t^4-m_H^4-2M_W^4-M_Z^4 \nonumber \\
&
\hspace{.5cm} +12 m_t^4 \ln\frac{\mu^2}{m_t^2} - m_H^4 \ln\frac{\mu^2}{m_H^2} 
 -6 M_W^4 \ln\frac{\mu^2}{M_W^2}  -3 M_Z^4 \ln\frac{\mu^2}{M_Z^2} 
\bigg] \,.
\end{align}

It is instructive to compare the threshold expansion for $g_3$ in terms of $\overline{g}_s$ with that in terms of
$g_s$. To do so, we first modify the threshold expansion in  Eq.~(\ref{eq:threshold_expansion}) to
\begin{align}
\label{eq:delta_g3_bar}
g_3 &= \overline{g}_s\left[1 +v_\sigma^2 \delta \overline{g}_3^{(6,0)}    + \delta \overline{g}_3^{(4,1)} +\delta \overline{g}_3^{(6,1,\sigma)}  \right] \, ,
\end{align}
where the coefficients on the right-hand side are understood to be a function of $\overline{g}_s$. 
Explicit results for both threshold expansions can be constructed from the expressions given in
Appendix~\ref{sec:threshold_implicit}. The  tadpole contributions, which are large numerically, enter the one-loop
dimension-six coefficients in Eq.~(\ref{eq:delta_g3_bar}) but not in Eq.~(\ref{eq:threshold_expansion}),
because in the latter case they are cancelled by contributions within the decoupling constants.

The analytic expressions for the expansion parameters of the threshold corrections given in Eq.~\eqref{eq:threshold_expansion} are included as ancillary files with the arXiv submission of this paper along with an example Mathematica notebook for the numerical evaluation using \texttt{LoopTools} and employing the experimental input values given in Table~\ref{tab:inputs}.

\section{Numerical analysis}
\label{sec:numerics}

\begin{table}[t]
	\begin{center}
		\def\arraystretch{1.3}
		\begin{tabular}{|cccc|}
			\hline 
			$G_F$ & $1.166378  \times 10^{-5} $ GeV$^{-2}$ & $\alpha (M_Z)^{-1}$ & $128.917 $ \\ 
			\hline
			$M_Z$ & $91.1880 $~GeV & $M_W$ & $80.3692$~GeV \\
			$M_H$ & $125.20 $~GeV & $m_t$ & $172.56 $~GeV \\
			$\alpha_s (M_Z)$ & $0.1190 \pm 0.0009$ & &  \\
			\hline 
		\end{tabular} 
		\caption{\label{tab:inputs}{ Experimental input values~\cite{ParticleDataGroup:2024cfk}. 
For reference, the values of the vev derived from these inputs are $v_\mu = 246.220$~GeV and 
$v_\alpha = 243.235$~GeV. } }
	\end{center}
\end{table}

The threshold corrections receive contributions from both QCD and EW effects, which factorise at one-loop order.  
The structure of the EW corrections is relatively intricate, as it depends on the EW input scheme and on the specific parameter or Wilson coefficient under consideration. In general,  the dominant EW contributions arise from top-quark loops and are proportional to the quantity\begin{align}
\label{eq:rho41}
\frac{\Delta \rho_t^{(4,1)}}{v_\sigma^2} = \frac{3}{16\pi^2} \frac{m_t^2}{v_\sigma^2}\approx  1\% \, .
\end{align}
The size of these top-loop (and other) corrections can be enhanced by a factor $c_w^2/s_w^2$ when arising from
the renormalisation of the sine of the  Weinberg angle, in which case their magnitude is set instead by the quantity
\begin{align}
-\frac{\Delta r_t^{(4,1)}}{v_\sigma^2} = \frac{c_w^2}{s_w^2}\frac{\Delta \rho_t^{(4,1)}}{v_\sigma^2}\approx 3.5\% \, .
\end{align}
Finally,    the threshold corrections to the Higgs-potential parameters $\lambda,\mu_H^2$ contain uncanceled tadpole 
contributions, which scale as 
\begin{align}
\frac{\Delta \rho_t^{{\rm tad}(4,1)}}{v_\sigma^2} \approx \frac{m_t^2}{m_H^2} \Delta \rho_t^{(4,1)} \approx 2 \Delta \rho_t^{(4,1)} 
\end{align}
and thus receive an additional enhancement.

\subsection{Threshold corrections at the EW scale}
\label{sec:NumThresh}

In order to illustrate these generic statements, let us first evaluate $g_1$ in the threshold expansion using
the numerical inputs in Table~\ref{tab:inputs}.  In this case, the tree-level SM results in the two
schemes are
\begin{align}
\label{eq:g1tree}
g_{1}^{v_\alpha(4,0)} = \frac{\sqrt{4\pi \alpha(M_Z)}}{c_w} = 0.3542 \, , \qquad  g_1^{v_\mu(4,0)} = \frac{2 M_Z s_w}{v_\mu} = 0.3499 \, . 
\end{align}
At NLO in SMEFT, for $\mu= M_Z$ the result in the $v_\alpha$ scheme is 
\begin{align}
\frac{g_1^{v_\alpha}}{g_1^{v_\alpha(4,0)}} & = 1.009+
 v_\alpha^2 \bigg[-(1\times 1.06) C_{HB}-(0.25\times 1.07) C_{HD}+0.023 C_{\substack{Hu\\33}}  - 0.019 C^{(1)}_{\substack{Hq\\33}}   + \dots              \bigg] \, .
\end{align}
 Here and below, the $\dots$ refer to SMEFT contributions which are smaller than those listed when all Wilson coefficients are set equal to a common value, and  the contributions from Wilson coefficients contributing already at LO are written in the
 form $(c \times K_{\rm NLO} )\, C_i$, where $c$ is the LO contribution and $K_{\rm NLO}$ is a multiplicative factor induced by the NLO correction.
For the same settings the NLO SMEFT result in the $v_\mu$ scheme is
\begin{align}
\frac{g_1^{v_\mu}}{g_1^{v_\mu(4,0)}} & =  1.024 +
 v_\mu^2 \bigg[-(1.87 \times 1.03) C_{HWB}- (1.12 \times 1.05) C_{HD}  
\nonumber \\ &
- (1 \times 1.04) C_{HB} - (0.5 \times 1.10) \left(C^{(3)}_{\substack{Hl \\ 11}}+ C^{(3)}_{\substack{Hl \\ 22}}  \right)
+ (0.5 \times 1.08) C _{\substack{ll \\ 1221}} 
\nonumber \\ &
+0.100 C_{\substack{Hu\\33}}  - 0.085 C^{(1)}_{\substack{Hq\\33}}   +0.050 C^{(3)}_{\substack{Hq\\33}}  
  + \dots              \bigg] \, .
\end{align}
One sees that the generic size of the NLO corrections is in line with the discussion above. 
In particular, one can  check using the analytic results in Appendix~\ref{sec:threshold_corrections} 
that the dominant contributions arise from top loops, and can also understand that  the corrections in the $v_\mu$ scheme tend to be larger, due to the fact that the tree-level SM result in that scheme is proportional to $s_w$ while that in the $v_\alpha$ scheme is not.  

This latter statement is not true in general, and is in fact reversed for the case of $g_2$, where the 
tree-level results are
\begin{align}
\label{eq:g2tree}
g_{2}^{v_\alpha(4,0)} = \frac{\sqrt{4\pi \alpha(M_Z)}}{s_w} = 0.6608 \, , \qquad  g_2^{v_\mu(4,0)} = \frac{2 M_Z c_w}{v_\mu} = 0.6528 \, . 
\end{align}
Again for the same inputs, in this case the NLO SMEFT results are 
\begin{align}
\frac{g_2^{v_\alpha}}{g_2^{v_\alpha(4,0)}} & = 0.9818 +
 v_\alpha^2 \bigg[(1.87 \times  0.981) C_{HWB}- (1 \times 1.03) C_{HW}  
+ (0.87 \times 0.989) C_{HD}  
\nonumber \\ &
-0.079 C_{\substack{Hu\\33}}  + 0.068 C^{(1)}_{\substack{Hq\\33}}  - 0.042 C^{(3)}_{\substack{Hq\\33}}  
  + \dots              \bigg] \, ,
\end{align}
and
\begin{align}
\frac{g_2^{v_\mu}}{g_2^{v_\mu(4,0)}} & = 0.9972 +
 v_\mu^2 \bigg[-(1\times 1.02) C_{HW}- (0.5 \times 1.02) \left(C^{(3)}_{\substack{Hl \\ 11}}
 + C^{(3)}_{\substack{Hl \\ 22}}  \right)+  (0.5 \times 1.00) C _{\substack{ll \\ 1221}} 
 \nonumber \\ &
 + 0.011 C_{HWB} + \dots              \bigg]  \, , 
\end{align}
so that the contributions are generally smaller in the $v_\mu$ scheme.

As noted above, larger contributions arise when tadpoles do not cancel, as is the case for the parameters in the Higgs potential.
For example, let us consider the threshold expansion for $\lambda$.  The tree-level SM results are 
\begin{align}
\label{eq:lamtree}
\lambda^{v_\alpha(4,0)} = \frac{m_H^2}{2v_\alpha^2} = 0.1325 \, , \qquad \lambda^{v_\mu(4,0)} =  \frac{m_H^2}{2v_\mu^2} = 0.1293 \, ,
\end{align}
while in SMEFT, again for $\mu=M_Z$, one has
\begin{align}
\frac{\lambda^{v_\alpha}}{\lambda^{v_\alpha(4,0)}} & = 1.091  +
 v_\alpha^2 \bigg[ (11.3\times 1.05)C_H + (3.73\times 1.21)C_{HWB}+ (2.24\times 1.17)C_{HD}
 \nonumber \\ &
 -(2\times 1.00)C_{H\Box} - 0.377 C_{\substack{uH\\33}} - 0.159 C_{\substack{Hu\\33}} + 0.135 C^{(1)}_{\substack{Hq\\33}} 
 - 0.109 C^{(3)}_{\substack{Hq\\33}} 
  + \dots              \bigg] \, ,
\end{align}
and
\begin{align}
\frac{\lambda^{v_\mu}}{\lambda^{v_\mu(4,0)}} & = 1.119 +
 v_\mu^2 \bigg[ (11.6\times 1.02)C_H - (2\times 1.00)C_{H\Box} - (1\times 1.27) \left(C^{(3)}_{\substack{Hl \\ 11}}
 + C^{(3)}_{\substack{Hl \\ 22}}  \right)
 \nonumber \\ &
 +(1\times 1.24)C _{\substack{ll \\ 1221}} +(0.5\times 1.00)C_{HD}-0.373 C_{\substack{uH\\33}}  - 0.029 C^{(3)}_{\substack{Hq\\33}} 
  + \dots              \bigg] \, .
\end{align}
One thus sees that the NLO threshold corrections $\lambda$ are typically of order 10\% times numerical factors depending
on the Wilson coefficient (and also in the SM), although they arise solely from the EW sector.   The situation is similar
for the parameter $\mu_H^2$, where the NLO SMEFT results are
\begin{align}
\frac{2}{m_H^2}\left(\mu_H^2 \right)^{v_\alpha} & =  
1.097 + v_\alpha^2 \bigg[
(5.66\times 1.11)C_H - (2\times 1.05)C_{H\Box}+ (0.5\times 1.39)C_{HD}
 \nonumber \\ &
+0.362 C_{HWB} - 0.337 C_{\substack{uH\\33}}   
  +  \left(4.9 C_{HW}   +   0.38 C_{HB} \right)\times 10^{-3}
 \bigg] \, ,
\end{align}
and
\begin{align}
\frac{2}{m_H^2}\left(\mu_H^2 \right)^{v_\mu} & =  
1.095 + v_\mu^2 \bigg[
(5.80\times 1.04 )C_H - (2\times 1.02)C_{H\Box}+ (0.5\times 1.02)C_{HD}
 \nonumber \\ &
 - 0.333 C_{\substack{uH\\33}}  + 0.095 C _{\substack{ll \\ 1221}} - 0.095 \left(C^{(3)}_{\substack{Hl \\ 11}}
 + C^{(3)}_{\substack{Hl \\ 22}}  \right)
 \nonumber \\ &
  +  \left(4.8 C_{HW}+ 0.70 C_{HWB}    +   0.37 C_{HB} \right)\times 10^{-3}
 \bigg] \, .
\end{align}

Finally, we list results for the top-quark Yukawa and strong coupling constant, both of which receive QCD as well as EW 
contributions.  For the top-quark Yukawa, the tree-level SM result is
\begin{align}
\label{eq:YTtree}
y_t^{v_\alpha(4,0)} = \frac{\sqrt{2}m_t}{v_\alpha} =1.003  \, , \qquad  y_t^{v_\mu(4,0)} = \frac{\sqrt{2}m_t}{v_\mu} = 0.9911 \, . 
\end{align}
The SMEFT results are:
\begin{align}
\frac{y_t^{v_\alpha}}{ y_t^{v_\alpha(4,0)}}& =  
0.9701 + v_\alpha^2 \bigg[
(1.87\times 0.945)C_{HWB} + (0.870\times 0.955)C_{HD}
+(0.498 \times 1.06) C_{\substack{uH \\ 33}} \, ,
 \nonumber \\ &
-0.081 C_{\substack{Hu\\ 33}} + 0.068 C^{(1)}_{\substack{Hq\\ 33}} -0.059  C_{\substack{uG\\ 33}} -0.052 C^{(3)}_{\substack{Hq\\ 33}}+ \dots
 \bigg] \, ,
\end{align}
and 
\begin{align}
\frac{y_t^{v_\mu}}{ y_t^{v_\mu(4,0)}}& =  
0.9857 + v_\mu^2 \bigg[
(0.504\times 1.03)C_{\substack{uH \\ 33}} - (0.5\times 0.992) \left(C^{(3)}_{\substack{Hl \\ 11}}
 + C^{(3)}_{\substack{Hl \\ 22}}  \right)
 \nonumber \\ &
+(0.5\times 0.967) C _{\substack{ll \\ 1221}} -0.058  C_{\substack{uG\\ 33}}  + \dots
 \bigg] \, .
\end{align}
For $g_3$ we have at tree level
\begin{align}
\label{eq:g3tree}
g_{3}^{v_\alpha(4,0)} =  g_3^{v_\mu(4,0)} = \sqrt{4\pi\alpha_s(M_Z)} = 1.22 \, . 
\end{align}
In the $v_\alpha$ scheme, the NLO corrections are given by
\begin{align}
\frac{g_3^{v_\alpha}}{g_1^{v_\alpha(4,0)}} & =  0.9960 +
 v_\alpha^2 \bigg[-(1.00 \times 1.05) C_{HG}+  0.020 C_{\substack{uG\\33}}   \bigg] \, , 
 \end{align}
while in the $v_\mu$ scheme
\begin{align}
\frac{g_3^{v_\mu}}{g_1^{v_\mu(4,0)}} & =  0.9960 +
 v_\mu^2 \bigg[-(1.00 \times 1.01) C_{HG}+  0.020  C_{\substack{uG\\33}}   \bigg] \,.
 \end{align}

\subsection{Threshold corrections in RG analyses}
\label{sec:RG_analysis}

SMEFT can be used both within a bottom-up approach, where the Wilson coefficients
are fit to data at the electroweak scale $\mu_{\rm EW}$, or within a top-down approach, where they are determined from a UV model at a high scale $\Lambda$ through matching.  In either case,
the parameters $p^{\rm sym}(\mu_{\rm EW})$ are determined through threshold corrections, and 
the sets of parameters $\{p^{\rm sym}, C_i\}$ at widely separated renormalisation scales are connected through
an RG analysis.  In this section, we briefly examine how the NLO threshold corrections in SMEFT
calculated in this work affect such analyses.  

Let us first consider the bottom-up approach. In this case we assume that a certain set of Wilson 
coefficients $c_i  \in C_i$ take on non-zero values at 
the EW scale, and use the threshold corrections in Eq.~(\ref{eq:threshold_expansion})
to calculate $p_i^{\rm sym}(\mu_{\rm EW})$ within a given EW input scheme.
This requires numerical values of the broken-phase parameters, which 
are listed in Table~\ref{tab:inputs}. Within a fixed-order analysis and to leading-logarithmic accuracy,
the Wilson coefficients and symmetric-phase parameters at a high scale $\Lambda$ 
can be written in terms of their values at the EW scale through the relations:
\begin{align}
\label{eq:1loopFixed}
p_i^{\rm sym}(\Lambda) & = p_i^{\rm sym}(\muEW) + \dot{p}_i^{\rm sym}(\muEW) 
 \ln\left(\frac{\Lambda}{\muEW} \right)\, , \nonumber \\
C_i(\Lambda) & = c_i(\muEW)+  \dot{C}_i(\muEW)\ln\left(\frac{\Lambda}{\muEW}\right) \, ,
\end{align}
where for a given quantity $x$ we define $\dot{x}\equiv dx/d\ln\mu$.
The values of the $c_i(\muEW)$ are fixed initial conditions, while the scale dependence of
the dotted objects can be written as
\begin{align}
\label{eq:1loopRunning}
 \dot{p}_i^{\rm sym}(\muEW)  & \equiv   \dot{p}_i^{\rm sym}\big( p_j^{\rm sym}(\muEW) ,c_j(\muEW)\big) \, ,  \nonumber \\
  \dot{C}_i^{\rm sym}(\muEW)  & \equiv   \dot{C}_i^{\rm sym}\big( p_j^{\rm sym}(\muEW) ,c_j(\muEW)\big) \, ,
\end{align}
which is meant to emphasise that the quantities with index $i$ depend in general on all the parameters
and Wilson coefficients under consideration. 
The fixed-order relations in Eq.~(\ref{eq:1loopFixed}) break down when the logarithm $\ln(\Lambda/\muEW) \gg 1$, in which case one must solve the RG equations numerically in order to resum logarithms. But they illustrate
the important point that within the bottom-up approach the dependence of the Wilson coefficients
on threshold corrections appears only through their RG evolution and such loop effects are 
expected to be small, unless the aforementioned logarithm is very large.

Let us now consider two simple scenarios within this bottom-up approach, involving modified top-quark couplings.
First, we study the case where only a single Wilson coefficient is set to a non-zero value at 
$\muEW=M_Z$,  namely 
\begin{align}
\label{eq:scenario1}
c_i(M_Z)  = C_{\substack{uH\\33}}(M_Z) = \frac{1}{1\, {\rm TeV}^2} \,.
\end{align}
We then examine the effect of LO and NLO threshold corrections on the symmetric-phase
parameters and Wilson coefficients at the high scale $\Lambda = 1$~TeV.  To this end,
we use the exact numerical solutions to the RG equations, with the only approximation being
that we restrict ourselves to the set of Wilson coefficients generated through the one-loop
running as written in Eq.~(\ref{eq:1loopRunning}); for this particular example it requires
the set  $C_i \in \{C_H, C_{\substack{uH\\33}}\}$.  The relevant set of RG equations
for the Wilson coefficients is then 
\begin{align}
16\pi^2 \dot{C}_{H} & = \left(-\frac{9}{2}g_1^2 -\frac{27}{2}g_2^2+108\lambda +18y_t^2 \right)C_H
+\left(24\lambda y_t -24y_t^3\right)C_{\substack{uH\\33}} \, , \nonumber \\
16\pi^2 \dot{C}_{\substack{uH\\33}} & = \left(-\frac{35}{12}g_1^2 -\frac{27}{4}g_2^2-8g_3^2
+24\lambda+\frac{51}{2}y_t^2\right)  C_{\substack{uH\\33}} \, , 
\end{align}
while for the SM parameters one has
\begin{align}
\dot{P}_i &= \dot{P}_i^{(4,1)} \, ,  \qquad P_i \in\{g_1,g_2,g_3,\mu_H^2\} \, ,  \nonumber\\
\dot{\lambda} & =\dot{\lambda}^{(4,1)} + \frac{3\mu_H^2}{4\pi^2}\left(2C_H + y_t C_{\substack{uH\\33}}\right) \, ,\nonumber \\
\dot{y}_t & = \dot{y}_t^{(4,1)} + \frac{3\mu_H^2}{8\pi^2}  C_{\substack{uH\\33}}  \,,
\end{align}
where expressions for the dotted symmetric-phase parameters in the SM limit, $p_i^{{\rm sym}(4,1)}$, are given in Eq.~(\ref{eq:p_sym_dot}) below.

Within this set-up, the effect of the tree and loop-level dimension-six threshold corrections
compared to the pure SM results at the scale $\muEW$ can be read off from the equations
given in Section~\ref{sec:NumThresh}.  In particular, compared to NLO SM, the tree-level 
SMEFT corrections increase the top-quark Yukawa $y_t$ by about 3\%; the NLO SMEFT corrections
to $y_t$ are negligible, but they serve to decrease the parameters $\lambda$ and $\mu_H^2$  by about 2\%. 

Consider now the evolution of the symmetric-phase parameters to the scale $\mu_{\rm NP} = \Lambda = 1$~TeV
in the SM limit, where all Wilson coefficients are set to zero.
Taking the NLO SM results for $p_i^{\rm sym}$ at $\muEW=M_Z$ within the $v_\mu$ scheme 
as the initial condition, and using the one-loop RG equations, we find
\begin{align}
\frac{p^{{\rm SM},v_\mu}_i(\Lambda)}{p^{{\rm SM},v_\mu}_i(M_Z)}& = \{1.01,0.980,0.685,1.07,0.891,0.872 \} \, , 
\end{align}
where the list of numbers on the right-hand side corresponds to the parameters $p_i^{\rm sym}$ in 
the order written in Eq.~(\ref{eq:p_sym}). If we now include 
SMEFT corrections, using first the NLO SM plus tree-level SMEFT threshold corrections as the initial conditions
(labelling the result with a superscript ``SM+(6,0)") and with the full one-loop RGEs, we find
\begin{align}
\frac{p^{{\rm SM} + (6,0),v_\mu}_i(\Lambda)}{p^{{\rm SM},v_\mu}_i(\Lambda)}& = \{1.00,1.00,0.925,1.00,1.04,1.00 \} \, , 
\end{align}
and thus see that $\lambda(\Lambda)$ is most strongly affected.  Finally, obtaining the initial
conditions from the full NLO threshold corrections in SMEFT (labelling the result with a superscript ``SM+(6,0)+(6,1)"), we find
\begin{align}
\frac{p^{{\rm SM} + (6,0)+(6,1),v_\mu}_i(\Lambda)}{p^{{\rm SM}+(6,0),v_\mu}_i(\Lambda)}& = \{1.00,1.00,0.958,0.981,1.00,1.00 \} \, ,
\end{align}
and see that adding the NLO dimension-six corrections to the initial conditions decreases the values of $\lambda$ and $\mu_H^2$ at the scale $\Lambda$ by 4\% and 2\%, respectively, meaning that loop effects are roughly half as large as the tree-level ones.

The threshold corrections also alter the Wilson coefficients at $\Lambda$, since the SM parameters that determine their anomalous dimensions are themselves changed.  In 
the scenario above, it turns out that the NLO threshold corrections affect the results for $C_i(\Lambda)$ below the 1\%  level.  However, this result is rather sensitive to the identities and numerical
values of the Wilson coefficients $c_i(M_Z)$ which define the initial conditions.
For instance, let us modify the scenario in Eq.~(\ref{eq:scenario1}) 
to include a non-zero value also for the Wilson coefficient of the top-quark chromomagnetic dipole operator:
\begin{align}
\label{eq:scenario2}
c_i(M_Z)  = \left\{C_{\substack{uH\\33}}(M_Z)   = \frac{1}{\Lambda^2} \, , C_{\substack{uG\\33}}(M_Z) =\frac{1}{\left(0.75\Lambda\right)^2} \right\}   \, . 
\end{align}
The top-quark dipole operator coefficient appears in loop-level threshold corrections only.  
Repeating the analysis above, one finds that the values of the SM parameters at the high scale $\Lambda = 1$~TeV are  rather minimally affected by NLO threshold corrections, and while the same is true of most Wilson coefficients at that 
scale\footnote{In this scenario the one-loop running involves  $C_i\in\{C_{HG},C_H,C_{\substack{uX\\33}}\}$, with $X=B,G,H,W$.}, the value of $C_{\substack{uH\\33}}$ ($C_H$) is shifted by 3\% (-2\%) compared to the case where the dimension-six threshold corrections 
are treated only at tree level.

Let us now comment briefly on a top-down approach.  One possible implementation, 
taken for instance in \cite{Aebischer:2018bkb}, is to  fix a numerical pattern of high-scale Wilson coefficients $c_i(\Lambda)$, and then use an iterative RG analysis to find the set of $p_i^{\rm sym}(\muEW), p_i^{\rm sym}(\Lambda),C_i(\muEW)$ that is consistent both with the high-scale Wilson coefficients and the 
values of the broken phase parameters as determined by threshold corrections at $\muEW$. 
We have studied the analogue of the scenarios in Eqs.~(\ref{eq:scenario1},\ref{eq:scenario2})
within this approach (i.e. that where the numerical values of the Wilson coefficients are taken
to be at $\Lambda$ rather than $M_Z$), and found that the effect of NLO threshold corrections is similar 
to that in the bottom-up approach analyzed above. We note that in model-specific implementations of the 
top-down approach, the Wilson coefficients $c_i(\Lambda)$ are calculable functions of both the symmetric phase and
new physics parameters, so the threshold corrections affect the Wilson coefficients beyond the 
one-loop running in Eqs.~(\ref{eq:1loopFixed},\ref{eq:1loopRunning}) and can thus lead to larger effects which must 
be determined on a case-by-case basis.

 \FloatBarrier
\section{Conclusions}
\label{sec:conclusions}

Threshold corrections provide a link between 
experimentally measured broken-phase input parameters and $\overline{\text{MS}}$-renormalised 
symmetric-phase parameters, and are thus a key ingredient for RG analyses and matching computations in SMEFT.  
In this work we have presented in analytic form the complete set of one-loop threshold corrections in
dimension-six SMEFT in two EW input schemes.   As a by-product of the analysis, we obtained
the one-loop  running of the strong coupling constant in the full EW SMEFT as well as the decoupling constant arising when this coupling is defined in a theory where the top quark and heavy electroweak bosons are integrated out.  We have performed extensive consistency checks on our results, including gauge independence, cancellation of divergences, and consistency with the known SM limits.

While modest in many scenarios, the NLO threshold corrections can be enhanced---for example in the Higgs
potential, which is sensitive to tadpole corrections, or more generally through contributions
involving top-quark loops.   A brief numerical analysis in two benchmark scenarios involving modified top-quark 
couplings illustrates that loop-level  dimension-six threshold corrections can change the SM parameters at the electroweak scale 
by up to $\mathcal{O}(5\text{--}10\%)$, with similar shifts to SM parameters and Wilson coefficients at the TeV scale.  Beyond this, our results provide a complete, process-independent set of one-loop
SMEFT threshold corrections that can be readily implemented in existing
RG-running frameworks such as  \cite{Celis:2017hod, Aebischer:2018bkb, DiNoi:2022ejg} to improve the theoretical 
accuracy of both top-down and bottom-up SMEFT analyses.   

\section*{Acknowledgements}

AB thanks Peter Stangl for useful discussions on the implementation of the RG evolution in \texttt{wilson}. 
This research was supported by the Deutsche Forschungsgemeinschaft (DFG, German Research Foundation) under grant 396021762 - TRR 257.

 \appendix  
 
 \section{SMEFT results for threshold corrections}
 \label{sec:threshold_corrections}
 
 In this section we work out the details of the general strategy outlined in Section~\ref{sec:details}
 for calculating the SMEFT expansion coefficients of the threshold corrections as defined in 
 Eq.~(\ref{eq:threshold_expansion}). Implicit results for the threshold corrections as a function
 of the finite parts of broken-phase counterterms are given in Section~\ref{sec:threshold_implicit},
 and results for these finite counterterm shifts in terms of Passarino-Veltmann (PV) integrals
 are given in Section~\ref{sec:CTs_with_PVs}.  In addition, we have included with the arXiv
 submission of this paper electronic files for the threshold corrections in terms of PV integrals
 directly. 
 
 \subsection{Threshold corrections from finite parts of counterterms}
 \label{sec:threshold_implicit}

As a first ingredient in this procedure, we define the SMEFT expansions used to renormalise
the broken-theory parameters listed in Eq.~(\ref{eq:p_broken}).  
For the Higgs vev, we follow \cite{Biekotter:2023xle} and write 
\begin{align}
\frac{1}{v_{T,0}^2}=\frac{1}{v_\sigma^2}\left[1 - v_\sigma^2 \delta v_\sigma^{(6,0)} -
\frac{1}{v_\sigma^2}\Delta v_\sigma^{(4,1)} - \Delta v_\sigma^{(6,1)} \right] \, ,
\end{align}
where here and below we use the notation of Eq.~(\ref{eq:param_CT}) for relations connecting
bare parameters with the renormalised ones plus counterterms, and the notation of 
Eq.~(\ref{eq:threshold_expansion}) for the superscripts on SMEFT expansion coefficients 
in a given scheme. In the two schemes, the tree-level results read
\begin{align}
\label{eq:vev_60}
 \delta v_\alpha^{(6,0)} & = -2\frac{c_w}{s_w}\left[C_{HWB} + \frac{c_w}{4s_w}C_{HD}\right] \, , \nonumber\\
  \delta v_\mu^{(6,0)} & = C^{(3)}_{\substack{Hl \\ 11}}+C^{(3)}_{\substack{Hl \\ 22}}- C_{\substack{ll \\ 1221}} \,. 
\end{align}

 For the heavy EW
boson masses $m_{\rm EW} \in \{M_W,M_Z,m_H\}$ we write
\begin{align}
\label{eq:mEW_CT}
m_{{\rm EW},0} &= m_{\rm EW}\left[1 + \frac{1}{v_\sigma^2}\Delta m_{\rm EW}^{(4,1)} +
\left( \Delta m_{\rm EW}^{(6,1)} - \delta v_\sigma^{(6,0)} \Delta m_{\rm EW}^{(4,1)}  \right) \right]  \nonumber \\
& \equiv m_{\rm EW}\left[1 + \frac{1}{v_\sigma^2}\Delta m_{\rm EW}^{(4,1)} +
  \Delta m_{\rm EW}^{(6,1,\sigma)}\right]  
\,.
\end{align}
The counterterm for $m_t$ receives both QCD and EW corrections at NLO in 
SMEFT, which we make explicit by writing
\begin{align}
m_{t,0} &= m_{t}\left[1 + \frac{1}{v_\sigma^2}\Delta m_{t}^{(4,1)} + \Delta m_{t}^{(6,1,\sigma)} 
+ \Delta m_t^{\alpha_s(4,1)} + \Delta m_t^{\alpha_s(6,1)} \right]\, ,
\end{align}
where the definition of  $\Delta m_{t}^{(6,1,\sigma)}$ follows  the notation of Eq.~(\ref{eq:mEW_CT}).
The counterterm for $g_s$ also receives QCD and EW corrections, but the latter only appear 
in the dimension-six piece, so in this case we write
\begin{align}
g_{s,0} &= g_s\left[1 +  \Delta g_s^{(4,1)} + \Delta g_s^{(6,1,\sigma)}\right]\, .
\end{align}

In addition one needs the analogous relations between the bare and renormalised 
symmetric-phase parameters in Eq.~(\ref{eq:p_sym}) and SMEFT Wilson coefficients in order 
to check that the threshold corrections are finite when the dimensional regulator is set to zero.
As an example, which is necessary for the numerical analysis in Section~\ref{sec:RG_analysis},
we list the one-loop counterterms in the pure SM \cite{Machacek:1981ic}.  Writing
\begin{align}
p_{i,0}^{\rm sym} = p_i^{\rm sym} + \frac{1}{2\epsilon}\left[ \dot{p}_i^{\rm sym}\right]^{(4,1)} \, ,
\end{align}
the explicit results when all SM quark flavours contribute to the running are 
\begin{align}
\label{eq:p_sym_dot}
16\pi^2 \dot{g}_1^{(4,1)} & = \frac{41}{6}g_1^3 \, , \quad 16\pi^2 \dot{g}_2^{(4,1)} = -\frac{19}{6}g_2^3 \, ,
\quad 16\pi^2 \dot{g}_3 = - 7 g_3^3  \, , \nonumber \\
16\pi^2 \dot{\lambda}^{(4,1)} &= 3\lambda\left(-g_1^2-3g_2^2+4y_t^2+8\lambda    \right)+\frac{3}{8}g_1^4+\frac{9}{8}g_2^4+\frac{3}{4} g_1^2 g_2^2-6y_t^4 \, , \nonumber \\
16\pi^2 \dot{\left[\mu_H^2\right]}^{(4,1)} &= \mu_H^2\left(-\frac{3}{2}g_1^2-\frac{9}{2}g_2^2 + 12 \lambda + 6 y_t^2\right)\, , 
\nonumber \\
16\pi^2 \dot{y}_t^{(4,1)} &= y_t\left(-\frac{17}{12}g_1^2 -\frac{9}{4}g_2^2 - 8 g_3^2 +\frac{9}{2}y_t^2 \right) \,.
\end{align}
The dimension-six counterterms for the symmetric-phase parameters are given in \cite{Jenkins:2013zja}, and 
that for the Wilson coefficients is in \cite{Jenkins:2013zja,Jenkins:2013wua,Alonso:2013hga}, but we do not
list them here.  Instead, we limit ourselves to pointing out a subtlety, which is 
that the aforementioned counterterms have been calculated directly in the symmetric
phase of the theory, while to cancel potential divergences in the threshold corrections they need to 
be expressed in terms of broken-theory parameters. These counterterms first appear at one
loop and contain only polynomial dependence on the couplings, and it is a simple matter to 
shift them to the broken phase using the tree-level relations of Eq.~(\ref{eq:tree_relations}).

In any case, splitting the broken-phase counterterms into divergent and finite parts using
the notation of Eq.~(\ref{eq:finite_split}), we find a complete cancellation of divergent pieces
in the SMEFT expansion coefficients of the threshold corrections. The results can  be 
written as follows. For $g_1$, 
\begin{align}
\delta g_1^{(6,0,\sigma)} & = -\frac{1}{4s_w^2} C_{HD} - \frac{c_w}{s_w} C_{HWB} - C_{HB} 
-\frac{1}{2} \delta v_\sigma^{(6,0)}  \, ,  \nonumber \\
\delta g_1^{(4,1,\sigma)} & = \delta M_Z^{(4,1)}+ \delta s_w^{(4,1)} - \frac{1}{2} \delta v_\sigma^{(4,1)}
\, , \nonumber \\
\delta g_1^{(6,1,\sigma)} & = \delta M_Z^{(6,1)}+ \delta s_w^{(6,1)} 
- \frac{1}{2} \delta v_\sigma^{(6,1)}-\frac{1}{4} \delta v_\sigma^{(6,0)} \left(6\delta M_Z^{(4,1)}+ 6\delta s_w^{(4,1)} +\delta v_\sigma^{(4,1)} \right) \nonumber \\&
-\frac{1}{4s_w^2}\left( \delta M_Z^{(4,1)}  -  \delta s_w^{(4,1)} + \frac{1}{2}\delta v_\sigma^{(4,1)}\right)C_{HD}
\nonumber \\
& -\frac{c_w}{s_w}\left( \delta M_W^{(4,1)} +\frac{1}{2} \delta v_\sigma^{(4,1)}\ \right)C_{HWB} \nonumber \\
& -\left( \delta M_Z^{(4,1)}+ \delta s_w^{(4,1)} + \frac{1}{2} \delta v_\sigma^{(4,1)}
 \right)C_{HB} \, ,
\end{align} 
where 
\begin{align}
\delta s_w^{(i,1)} & = -\frac{c_w^2}{s_w^2}\left(\delta M_W^{(i,1)} -\delta M_Z^{(i,1)} \right) \,.
\end{align}
For $g_2$, we have 
\begin{align}
\delta g_2^{(6,0,\sigma)} & = - C_{HW} -\frac{1}{2} \delta v_\sigma^{(6,0)} \nonumber  \, , \\
\delta g_2^{(4,1,\sigma)} & = \delta M_W^{(4,1)} - \frac{1}{2} \delta v_\sigma^{(4,1)} \nonumber \, , \\
\delta g_2^{(6,1,\sigma)} & = \delta M_W^{(6,1)} - \frac{1}{2} \delta v_\sigma^{(6,1)} 
- \frac{1}{2} \delta v_\sigma^{(6,0)} \left(3\delta M_W^{(4,1)} +\frac{1}{2}\delta v_\sigma^{(4,1)} \right) \nonumber \\
& -\left(\delta M_W^{(4,1)} + \frac{1}{2}\delta v_\sigma^{(4,1)}  \right)C_{HW} \,.
\end{align}
For $\lambda$ we find 
\begin{align}
\delta \lambda^{(6,0,\sigma)} & = \frac{3 v_\sigma^2}{m_H^2}  C_{H} - 2 C_{H,{\rm kin}}
- \delta v_\sigma^{(6,0)} \nonumber \, , \\
\delta \lambda^{(4,1,\sigma)} & =2 \delta m_H^{(4,1)} - \delta v_\sigma^{(4,1)} \nonumber \, ,\\
\delta \lambda^{(6,1,\sigma)} & = 2\delta m_H^{(6,1)} - \delta v_\sigma^{(6,1)} 
-4 \delta v_\sigma^{(6,0)}   \delta m_H^{(4,1)}+ \frac{3 v_\sigma^2}{m_H^2}  \delta v_\sigma^{(4,1)} C_{H} -4\delta m_H^{(4,1)}  C_{H,{\rm kin}} \, , 
\end{align}
where $C_{H,{\rm kin}}  \equiv C_{H\Box}- C_{HD}/4$.
For $\mu_H^2$ we find
\begin{align}
 \left[\delta\mu_H^2\right]^{(6,0,\sigma)} & = \frac{3 v_\sigma^2}{2m_H^2}  C_{H} - 2 C_{H,{\rm kin}} \, , \nonumber \\
 \left[\delta\mu_H^2\right]^{(4,1,\sigma)} & =2 \delta m_H^{(4,1)} \nonumber  \, ,\\
 \left[\delta\mu_H^2\right]^{(6,1,\sigma)} &  = \delta \lambda^{(6,1,\sigma)}+\delta v_\sigma^{(6,1)}
+2 \delta v_\sigma^{(6,0)}  \delta m_H^{(4,1)}
-2 \delta v_\sigma^{(4,1)}C_{H,{\rm kin}}  \, .
\end{align}
Regarding the top-quark Yukawa coupling $y_t$, the EW corrections read
\begin{align}
\delta y_t^{(6,0,\sigma)} & = \frac{v_\sigma}{2\sqrt{2} m_t} C_{\substack{uH\\33}} 
  - \frac{1}{2} \delta v_\sigma^{(6,0)} \nonumber \, , \\
\delta y_t^{(4,1,\sigma)} & = \delta m_t^{(4,1)} - \frac{1}{2} \delta v_\sigma^{(4,1)} \nonumber \, ,\\
\delta y_t^{(6,1,\sigma)} & = \delta m_t^{(6,1)} - \frac{1}{2}\delta v_\sigma^{(6,1)} 
 -\frac{1}{2}\left(3 \delta m_t^{(4,1)} - \frac{1}{2} \delta v_\sigma^{(4,1)}\right)\delta v_\sigma^{(6,0)}
 +\delta y_t^{(6,0,\sigma)} \delta v_\sigma^{(4,1)}
   \, , 
\end{align}
while the QCD ones are given by
\begin{align}
\label{eq:mt_qcd}
\delta y_t^{\alpha_s(4,1)} & = \delta m_t^{\alpha_s(4,1)} = 
-\frac{C_F\alpha_s}{\pi}\left(1+ \frac{3}{4}\ln \frac{\mu^2}{m_t^2}\right) \, ,
\nonumber \\
\delta y_t^{\alpha_s(6,1,\sigma)} & = \delta m_t^{\alpha_s(6,1)} -\frac{1}{2}v_\sigma^2 \delta m_t^{\alpha_s(4,1)} \delta v_\sigma^{(6,0)}  \nonumber \\
& =  \frac{C_F g_s}{4\pi^2}\frac{m_t v_\sigma}{\sqrt{2}}\left(1 +3 \ln \frac{\mu^2}{m_t^2}\right) C_{\substack{uG\\33}} -\frac{1}{2}v_\sigma^2 \delta m_t^{\alpha_s(4,1)} \delta v_\sigma^{(6,0)}   \,.
\end{align}

Finally, for the threshold corrections in Eq.~(\ref{eq:delta_g3_bar}) relating $g_3$ with $\overline{g}_s$ in the full EW SMEFT, which include contributions from the top quark and heavy weak bosons in the running, we have
\begin{align}
\label{eq:delta_g3_bar_coeffs}
\delta \overline{g}_3^{(6,0)} & = -  C_{HG} \, , \nonumber \\
\delta \overline{g}_3^{(4,1)} & =  0  \, , \nonumber \\
\delta \overline{g}_3^{(6,1,\sigma)} & =  - \delta v_\sigma^{(4,1)}C_{HG}\,.
\end{align}
Those relating  $g_3$ with $g_s$ defined in Eq.~(\ref{eq:threshold_expansion}) make use of the decoupling constants in
Eq.~(\ref{eq:alpha_decoup}) and are given by 
\begin{align}
\delta g_3^{(6,0)}& =\delta \overline{g}_3^{(6,0)}  \, , \nonumber \\
\delta g_3^{(4,1 )}& = \frac{1}{2}\zeta_{\alpha_s}^{(4,1)}  \, , \nonumber \\
\delta g_3^{(6,1,\sigma)}& = \delta \overline{g}_3^{(6,1,\sigma)} +\frac{1}{2}\zeta_{\alpha_s}^{(6,1,\sigma)} 
+\frac{1}{2}\zeta_{\alpha_s}^{(4,1)} \delta \overline{g}_3^{(6,0)} v_\sigma^2
\,.
\end{align}

 \subsection{Finite parts of counterterms as a function of PV integrals}
 \label{sec:CTs_with_PVs}

In this section we list the finite parts of the broken-phase counterterms needed to evaluate the threshold
corrections explicitly.  

We begin by introducing some shorthand notation.  For squared masses, we use 
\begin{align}
\label{eq:mass_shorthand}
h=m_H^2, \, w = M_W^2, \,  z = M_Z^2, \, t=m_t^2 \, . 
\end{align}
All results can be written in terms of the single loop function
\begin{align}
B_0\left(s; m_1^2,m_2^2\right)= -\int_0^1 dx \ln\left(\frac{x m_1^2+(1-x)m_2^2 - x(1-x)s}{\mu^2}\right) \,,
\end{align}
which is the finite part of a bubble integral.  We also express tadpole integrals in terms of this
function, defining
\begin{align}
B_0\left(0; 0, x \right)& \equiv B_{x} = 1 - \ln\left(\frac{x}{\mu^2}\right)\, , 
\end{align}
where $x$ can be any of the squared masses in Eq.~(\ref{eq:mass_shorthand}).  We shall
make use of the following tadpole functions in unitary gauge
\begin{align}
\label{eq:Tfuns}
  T^{(4,1)} & = -4w^2-2z^2+\frac{3}{2}h^2 B_h-12 t^2 B_t+6 w^2 B_w + 3 z^2 B_z \, , 
\nonumber \\
T^{(6,1)} & =  hB_h\left( -3 C_H v_\sigma^2 +2 h C_{H,{\rm kin}}\right)
+w^2\left(-8 + 12 B_w\right)C_{HW} 
\nonumber \\ &
+ z^2\left(-1+\frac{3}{2}B_z \right)\left(C_{HD} + 4 c_{hzz}\right)
+6\sqrt{2}m_t v_\sigma t B_t C_{tH} + C_{H,{\rm kin}} T^{(4,1)} \,,
\end{align}
where we have defined the following combinations of Wilson coefficients:
\begin{align}
C_{H,{\rm kin}} = C_{H\Box} -\frac{1}{4}C_{HD} \, , \quad 
c_{hzz}= c_w^2 C_{HW}+c_w s_w C_{HWB}+s_w^2 C_{HB} \,.
\end{align}
Finally, to avoid a proliferation of flavour indices it is convenient to introduce the following abbreviations 
for (sums) of Wilson coefficients,
\begin{align}
C_{Hl}& \equiv \sum_{i=1}^{3}  C_{\substack{Hl\\ii}}, \, \quad 
C_{Hq}  \equiv \sum_{j=1}^{2}  C_{\substack{Hq\\jj}} \ , \quad
C_{HQ} \equiv  C_{\substack{Hq\\33}}   \,, \nonumber \\
C_{Hd}& \equiv \sum_{i=1}^3  C_{\substack{Hd\\ii}} \, , \quad C_{He} \equiv \sum_{i=1}^3  C_{\substack{He\\ii}} \, ,\quad C_{Hu} \equiv \sum_{i=1}^2  C_{\substack{Hu\\jj}}  \, , \nonumber \\ 
C_{Ht} &  \equiv   C_{\substack{Hu\\33}} \, , \quad C_{tB}  \equiv  C_{\substack{uB\\33}} 
 \, , \quad C_{tW}  \equiv   C_{\substack{uW\\33}}  \, , \quad C_{tH}  \equiv   C_{\substack{uH\\33}}   \,,
\end{align}
the last of which was already used in Eq.~(\ref{eq:Tfuns}). It is understood that the notation for $C_{Hl}$ applies 
to both $C^{(1)}_{Hl}$ and $C^{(3)}_{Hl}$, and similarly for $C_{Hq}$ and $C_{HQ}$.

With this notation, the SM part of the Higgs counterterm is 
\begin{align}
16\pi^2 \delta m_H^{(4,1)} & =
-6\frac{w^2}{h} -3 \frac{z^2}{h}+\frac{3}{4} hB_h -6 \frac{t^2}{h}B_t +
\left(-w+3 \frac{w^2}{h}\right)B_w 
+\left(\frac{3}{2}\frac{z^2}{h}-\frac{z}{2}\right)B_z
 \nonumber \\ &
 +\frac{9}{4}h B_0(h;h,h) + \left(3 t - 12 \frac{t^2}{h} \right)B_0(h; t,t)
 +\left(\frac{h}{2} - 2 w +6 \frac{w^2}{h}\right)B_0(h; w, w)
  \nonumber \\ &
   +\left(\frac{h}{4} + 3\frac{z^2}{h} -z \right)B_0(h; z,z)  - \frac{3}{2h} T^{(4,1)}\, ,
\end{align}
while the SMEFT result reads

 \begin{align}
16\pi^2 &\delta m_H^{(6,1)} = -12\frac{w^2}{h}C_{H,{\rm kin}}-\frac{3z^2}{h}\left(2 C_{H\Box}+C_{HD}\right) + \left(8w - 36\frac{w^2}{h} \right)C_{HW}
\nonumber \\ &
+ \left(4z - 18 \frac{z^2}{h}\right)c_{hzz}
-9 v_\sigma^2\left(B_h +  B_0(h;h,h)\right)C_{H} + \frac{h}{2}\left(19B_h + 27 B_0(h;h,h)\right)C_{H,{\rm kin}}
\nonumber \\ &
+\left[\left(-2w+6\frac{w^2}{h}\right)B_w + \left(h-4w+12\frac{w^2}{h}\right)B_0(h;w,w)    \right]C_{H,{\rm kin}} 
\nonumber \\ &
+\left[30\frac{w^2}{h}B_w + \left(-12w +24\frac{w^2}{h}\right)B_0(h;w,w)    \right]C_{HW} 
+\frac{9z^2}{4h}B_z C_{HD}
\nonumber \\ &
+\left[\left(-z+3\frac{z^2}{h}\right) B_z +\left(\frac{h}{2}-2z + 6\frac{z^2}{h}\right)B_0(h;z,z) \right]\left(C_{H\Box}+\frac{1}{4}C_{HD}\right)
\nonumber \\ &
+\left[ \frac{15z^2}{h}  B_z +\left(-6z + 12\frac{z^2}{h}\right)B_0(h;z,z) \right]c_{hzz}
\nonumber \\ &
+\left[-12\frac{t^2}{h}B_t + \left(6t-24\frac{t^2}{h}\right)B_0(h;t,t) \right]C_{H,{\rm kin}}
\nonumber \\ &
+\sqrt{2}m_ tv_\sigma\left[15\frac{t}{h} B_t + \left(-3+12\frac{t}{h}\right)B_0(h;t,t) \right]C_{tH}
\nonumber \\ &
-\frac{1}{h}\left[T^{(4,1)}\left(-3\frac{v_\sigma^2}{m_H^2}C_H + \frac{7}{2}C_{H,{\rm kin}}\right)
+\frac{3}{2}T^{(6,1)} \right]\,.
\end{align}
The corresponding results for the $W$-boson mass are
 \begin{align}
16\pi^2 \delta m_W^{(4,1)} & =
-\frac{h}{3}+2t -\frac{98}{9}w +8 w c_w^2 -\frac{z}{3}
+\left(\frac{h}{2}-\frac{h^2}{6w}\right)B_h
+\left(-2t+\frac{t^2}{w}\right)B_t 
 \nonumber \\ &
 +\left(\frac{h}{6}+8 w c_w^2 +\frac{z}{6} \right)B_w
 +\left(4w-\frac{4z}{3}  -\frac{z}{6c_w^2} \right)B_z
  +\left( -t -\frac{t^2}{w}+2w \right)B_0(w;0,t)
   \nonumber \\ &
   +\left(-\frac{2h}{3} + \frac{h^2}{6w}+2w\right)B_0(w;h,w)
   +\left(-\frac{34w}{3} - 8 w c_w^2 +\frac{8z}{3} + \frac{z}{6c_w^2} \right)B_0(w;w,z) 
      \nonumber \\ &
      -\frac{1}{h}T^{(4,1)} \, ,
\end{align}
and
\begin{align}
16\pi^2 &\delta m_W^{(6,1)}=
-\frac{2h}{3}C_{H,{\rm kin}}+ 4 t C_{HQ}^{(3)} + \frac{4w^2}{z}C_{HD}
+z\left(-\frac{1}{6}C_{HD}-\frac{20c_ws_w}{3}C_{HWB}\right)
\nonumber \\ &
 + w \left(-\frac{4}{9} C_{H\Box} -\frac{4}{3}C_{HD}-\frac{4}{3}C_{HQ}^{(3)}+16 c_w s_w C_{HWB}  
 +20\frac{M_W}{v_\sigma}C_W + \frac{8}{9}C_{Hl}^{(3)} + \frac{8}{3}C_{Hq}^{(3)}\right) 
 \nonumber \\ &
 +h\left(C_{H,{\rm kin}}+ 5 C_{HW}-\frac{h}{3w} C_{H,{\rm kin}} \right)B_h
 +t \left( -4 C_{HQ}^{(3)} -6\sqrt{2}\frac{m_t}{M_W}C_{tW}  +\frac{2t}{w}     C_{HQ}^{(3)}    \right)B_t
  \nonumber \\ &
  +\bigg[w\left(-\frac{2}{3}C_{H\Box}+\frac{5}{3}C_{HD}+\frac{4}{3}C_{Hl}^{(3)}+4 C_{Hq}^{(3)}  -4C_{HW}+16 c_w s_w C_{HWB} -12\frac{M_W}{v_\sigma}C_W \right) 
    \nonumber \\ &
 +\frac{h}{3}C_{H,{\rm kin}}+\frac{4w^2}{z}C_{HD}+z\left(\frac{1}{12}C_{HD}+\frac{10c_ws_w}{3}C_{HWB}\right)    \bigg] B_w
      +  \bigg[ 2w C_{HD} 
    \nonumber \\ &+ z\left(-\frac{2}{3}C_{HD}+8 c_w s_w C_{HWB} -12 \frac{M_W}{v_\sigma}C_W\right) -\frac{z^2}{w}\left(\frac{1}{12}C_{HD}+\frac{10c_ws_w}{3}C_{HWB}\right)\bigg]B_z
     \nonumber \\ &
   +t  \left[\left(-2-\frac{2t}{w}+\frac{4w}{t}\right)C_{HQ}^{(3)} - 6\sqrt{2}\left(\frac{M_W}{m_t}-\frac{m_t}{M_W}\right)C_{tW}\right]B_0(w;0,t)
       \nonumber \\ &
     +  \bigg[-h\left(\frac{4}{3}C_{H,{\rm kin}}+4C_{HW}\right)+\frac{h^2}{3w}C_{H,{\rm kin}}
       +4w\left(C_{H,{\rm kin}}+ 2 C_{HW}\right)\bigg]B_0(w;h,w)
          \nonumber \\ &
        +\bigg[w\left(-\frac{17}{3}C_{HD}-16 c_w s_w C_{HWB} - 48\frac{M_W}{v_\sigma}C_W\right) 
        -\frac{4w^2}{z}C_{HD}
          \nonumber \\ &
          +z\left(\frac{4}{3}C_{HD}-\frac{28}{3}c_w s_w C_{HWB} + 12\frac{M_W}{v_\sigma}C_W\right)
           \nonumber \\ &
           +\frac{z^2}{w}\left(\frac{1}{12}C_{HD}+\frac{10}{3}c_w s_w C_{HWB}\right)   \bigg]B_0(w;w,z) 
           \nonumber \\ &
     -\frac{1}{h}\left(T^{(4,1)}\left(C_{H,{\rm kin}}+2 C_{HW}\right) + T^{(6,1)} \right)      \,.
\end{align}
While the SM result for the $Z$-boson mass counterterm is relatively simple,  
\begin{align}
16\pi^2 \delta m_Z^{(4,1)} & =
-\frac{h}{3}+\frac{34t}{9}-\frac{130w}{9}+ \frac{64}{9} \frac{tw^2}{z^2}- \frac{8w^3}{z^2}
-\frac{80}{9} \frac{tw}{z} + \frac{40}{3}\frac{w^2}{z}+\frac{62 z}{9}
      \nonumber \\ &
      +\left(\frac{h}{2}-\frac{h^2}{6z} \right)B_h
      +\left(-\frac{34t}{9}-\frac{64}{9}\frac{tw^2}{z^2}+\frac{80}{9}\frac{tw}{z} \right)B_t
      +\left(-\frac{w}{3}+\frac{8w^3}{z^2} - \frac{8w^2}{3z}\right)B_w
          \nonumber \\ &
          +\left(\frac{h}{6}- \frac{200w}{9}+\frac{160}{9}\frac{w^2}{z}+\frac{100z}{9}\right)B_z
          +\left(-\frac{2h}{3}+\frac{h^2}{6z}+2z \right)B_0(z;h,z)
           \nonumber \\ &
          +\frac{1}{9} \left(7t-40w+64\frac{tw^2}{z^2}-80 \frac{tw}{z}+32 \frac{w^2}{z}+17z \right)B_0(z;t,t)
            \nonumber \\ &
        +    \left(\frac{8w}{3}-\frac{8w^3}{z^2}-\frac{34w^2}{3z}+\frac{z}{6} \right)B_0(z;w,w) - \frac{1}{h}T^{(4,1)} \,,
\end{align}
the dimension-six part receives contributions from a large number of fermion-specific $Z$-boson couplings and 
requires some additional notation.  To this end, we first split up the result as 
\begin{align}
 \delta m_Z^{(6,1)}= \delta m_{Z,{\rm fer}}^{(6,1)} + \delta m_{Z,{\rm bos}}^{(6,1)} 
  + \delta m_{Z,{\rm tad}}^{(6,1)}  \,, 
\end{align}
where the tadpole contributions read
\begin{align}
16\pi^2 &\delta m_{Z,{\rm tad}}^{(6,1)}= -\frac{1}{h}\left(T^{(4,1)}\left(C_{H\Box}+\frac{1}{4}C_{HD}
+ 2 c_{hzz}\right)+ T^{(6,1)}\right) \, , 
\end{align}
and the one-particle irreducible (1PI) bosonic contributions are given by
\begin{align}
16\pi^2 &\delta m_{Z,{\rm bos}}^{(6,1)}= \left(-\frac{5w}{9}-\frac{4w^3}{z^2}+\frac{4w^2}{3z}+\frac{z}{18}\right)C_{HD}+\frac{s_w}{c_w}\left(-\frac{4w}{9}- 16\frac{w^3}{z^2}+\frac{8w^2}{3z}\right)C_{HWB}
\nonumber \\
& + \left(-\frac{2h}{3}-\frac{4z}{9}\right)\left(C_{H\Box}+\frac{1}{4}C_{HD}\right)
\nonumber \\ &
+h\left[C_{H\Box}+C_{HD}+5 c_{hzz}-\frac{h}{3z}\left(C_{H\Box}+\frac{1}{4}C_{HD}\right)\right]B_h
\nonumber \\ &
+w\left[\left(-\frac{1}{6}+ \frac{4w^2}{z^2} - \frac{4w}{3z}\right)C_{HD} 
+\frac{s_w}{c_w}\left(16\frac{w^2}{z^2}-\frac{20w}{3z}\right)C_{HWB}\right]B_w
\nonumber \\ &
+z\left[\left(-\frac{2}{3}+ \frac{h}{3z}\right)\left(C_{H\Box}+\frac{1}{4}C_{HD}\right)    -4 c_{hzz}  \right]B_z
\nonumber \\ &
+z\left[\left(4+\frac{h^2}{3z^2}-\frac{4h}{3z}\right)\left(C_{H\Box}+\frac{1}{4}C_{HD}\right)   
+\left(8-\frac{4h}{z}\right)c_{hzz}  \right]B_0(z;h,z)
\nonumber \\ &
+z\bigg[\left(\frac{1}{12}-\frac{4w^3}{z^3}-\frac{17}{3}\frac{w^2}{z^2}+\frac{4w}{3z} \right)C_{HD}  
\nonumber \\ &
\hspace{1cm}
+\frac{s_w}{c_w}\left( -16\frac{w^3}{z^3}-\frac{28}{3}\frac{w^2}{z^2}+
\frac{10w}{3z}  \right)C_{HWB}\bigg]B_0(z;w,w)
\nonumber \\ &
-z\frac{M_W}{v_\sigma}\left[\frac{4w}{z}-24\frac{w^2}{z^2}+24\frac{w^2}{z^2}B_w
+\left(48\frac{w^2}{z^2}-12\frac{w}{z}\right)B_0(z;w,w)
\right]C_W \, .
\end{align}
The 1PI fermionic contribution is 
 \begin{align}
 16\pi^2 &\delta m_{Z,{\rm fer}}^{(6,1)}= \frac{z}{9}\left(2+3B_z\right)\left[F^{\ell}+F^{\nu}+3 F^u + 3 F^d\right] \nonumber \\
 & - 
 2\sqrt{2}m_t M_Z\left(5-\frac{8w}{z}\right)\left(s_w C_{tB}-c_w C_{tW}\right)B_0(z;t,t)
\nonumber \\ 
&+  z\left[  -\frac{1}{3}+\frac{2t}{z}-\frac{2t}{z}B_t+\left(1-\frac{t}{z}\right)B_0(z;t,t)  \right] F'   \nonumber \\
&+6t\left[{\cal Z}_L^{4u}Z_R^{6t}  +{\cal Z}_R^{4u}Z_L^{6t} -
\frac{1}{2} {\cal Z}_L^{4u} {\cal Z}_R^{4u}C_{HD} \right]B_0(z;t,t)    \, .
\end{align}
The contributions from light-fermions $f\in \{\ell,\nu,d,u\}$ can be written as 
 \begin{align}
 F^f = 2 {\cal Z}_L^{4f}{\cal Z}_L^{6f}+2 {\cal Z}_R^{4f}{\cal Z}_R^{6f} 
 -\frac{1}{2}\left(3-\delta_{fu}\right)\left[ \left({\cal Z}_L^{4f}\right)^2+ \left({\cal Z}_R^{4f} \right)^2\right]C_{HD} \, , 
 \end{align}
where $\delta_{fu}=1$ for $f=u$ and vanishes for all other $f$. This result involves
the left- and right-handed $Z$-fermion couplings in the SM, 
\begin{align}
\label{eq:ZLR4}
 {\cal Z}_L^{4f}  = 2 T_3^f - 2 s_w^2 Q_f \, , \qquad  {\cal Z}_R^{4f}  =   - 2 s_w^2 Q_f \,,
\end{align}
where $Q_f$ is the charge and $T_3^f$ is the third component of the  weak isospin of fermion $f$,
as well as the corresponding sums of contributions in SMEFT, which can be 
written as 
\begin{align}
 {\cal Z}_{L/R}^{6f} & =\left(3-\delta_{fu}\right)Q_f\left(c_w^2 C_{HD}+2 c_w s_w C_{HWB}\right)  +  g^{6f}_{L/R}  \,.
\end{align}
The flavour-specific results are 
 \begin{align}
	\label{eq:6f}
 g^{6\ell}_L & = -   C_{Hl}^{(1)}  - C_{Hl}^{(3)}\, ,   &
g^{6\ell}_R  &= -   C_{He } \,, \nonumber \\
 g^{6\nu}_L & = -   C_{Hl}^{(1)}  + C_{Hl}^{(3)}\, ,   &
g^{6\nu}_R & =0 \, ,\nonumber \\
g^{6u}_L & = -   C_{Hq }^{(1)}  + C_{Hq}^{(3)}\, ,   &
g^{6u}_R & = -   C_{Hu} \ , \nonumber \\
g^{6d}_L & = -   C_{Hq }^{(1)}  - C_{Hq}^{(3)}-   C_{HQ }^{(1)}  - C_{HQ}^{(3)}\, ,  &
g^{6d}_R  &=  - C_{Hd} \,.
\end{align} 
The contributions from the top quark in addition involve the functions
\begin{align}
 Z_{L}^{6t} & =Q_u\left(c_w^2 C_{HD}+2 c_w s_w C_{HWB}\right)  -C_{HQ}^{(1)}+C_{HQ}^{(3)}  \,,
 \nonumber \\
  Z_{R}^{6t} & =Q_u\left(c_w^2 C_{HD}+2 c_w s_w C_{HWB}\right) -C_{Ht}\, , 
\end{align}
and
\begin{align}
\label{eq:ZLZR_6}
 F' = 2 {\cal Z}_L^{4u}Z_L^{6t}+2 {\cal Z}_R^{4u}Z_R^{6t} 
 -\frac{1}{2}\left[ \left({\cal Z}_L^{4u}\right)^2+ \left({\cal Z}_R^{4u} \right)^2\right]C_{HD} \, .
 \end{align}

The QCD counterterms for the top-quark mass have been given in Eq.~(\ref{eq:mt_qcd}),  while for the EW ones
we draw on the notation above in writing the pieces involving top-$Z$ couplings.  In particular, the SM result is
\begin{align}
 16\pi^2  \delta m_t^{(4,1)} & =  \frac{16}{9}M_W^2 s_w^2 f^{\gamma} +  2 w f^{w} + t f^h + z\left( \left({\cal Z}_L^{4u}\right)^2+ \left({\cal Z}_R^{4u} \right)^2\right)f_1^z + z {\cal Z}_L^{4u}{\cal Z}_R^{4u}f_2^z
  \nonumber \\ &
-\frac{1}{h}T^{(4,1)}  \, ,
\end{align}
where  
\begin{align}
\label{eq:ft_funs}
f^\gamma &  = -1-3B_t \, ,
\nonumber \\
f^w & = -\frac{1}{2} + \left( -\frac{1}{4}+\frac{w}{2t}\right)B_w + \left(\frac{1}{4}+\frac{t}{4w}-\frac{w}{2t}\right)B_0(t;0,w) \, , 
\nonumber \\
f^h& = \frac{h}{2t}B_h - \frac{1}{2}B_t + \left(2-\frac{h}{2t}\right)B_0(t;h,t) \, , \nonumber \\
  f_1^z & = \frac{1}{2} \left\{ -1-\left(1+\frac{t}{z}\right)B_t  +\frac{z}{t}B_z 
  + \left(1-\frac{z}{t}\right)B_0(t;t,z)  \right\}\, ,
\nonumber \\
f_2^z & = 2 +\frac{t}{z}B_t -3 B_0(t;t,z) \,,
\end{align}
while in SMEFT
\begin{align}
16\pi^2 &\delta m_t^{(6,1)}  =  \frac{16}{9}M_W^2 s_w^2 f^{\gamma}\delta v_\alpha^{(6,0)}
+ 4wf^w C_{HQ}^{(3)}+2t\left(C_{H, {\rm kin}}-\frac{1}{\sqrt{2}} \frac{v_\sigma}{m_t}C_{tH}  \right)f^h  
\nonumber \\ &
+z F' f_{1}^z+z\left({\cal Z}_L^{4u}Z_R^{6t}  +{\cal Z}_R^{4u}Z_L^{6t} -
\frac{1}{2} {\cal Z}_L^{4u} {\cal Z}_R^{4u}C_{HD}\right)f_{2}^z
\nonumber \\ 
 & -\frac{3v_\sigma}{2\sqrt{2}m_t} h B_h C_{tH} 
 - \frac{ \sqrt{2} M_W}{m_t}\left(-2t+4w - 3w B_w + 3(t-w)B_0(t;0,w)  \right) C_{tW}
 \nonumber \\ & 
 +\frac{\sqrt{2}M_Z}{m_t}\left(  {\cal Z}_L^{4u} + {\cal Z}_R^{4u} \right)
 \left(2(t+z)-3t B_t -3 z B_0(t;t,z)\right)\left(s_w C_{tB} - c_w C_{tW} \right)
 \nonumber \\ & 
 - \sqrt{2}m_t M_W s_w\left(\frac{16}{3}- 8B_t\right)\left(c_w C_{tB} + s_w C_{tW} \right)
 +t\left(2-4B_t\right)  \left(C^{(1)}_{\substack{qu\\3333}} + \frac{4}{3} C^{(8)}_{\substack{qu\\3333}} \right) 
 \nonumber \\ &
-\frac{1}{h}\left(T^{(6,1)} + \left(C_{H, {\rm kin}}-\frac{1}{\sqrt{2}} \frac{v_\sigma}{m_t}C_{tH}  \right)T^{(4,1)} \right)\, .
\end{align}

Finally, the tree-level vev shifts in both schemes have been given in Eq.~(\ref{eq:vev_60}).  In the $v_\mu$
scheme, the one-loop SM result is
\begin{align}
16\pi^2 \delta v_\mu^{(4,1)} & = -\frac{h}{2}+3t - w -\frac{z}{2}+ \frac{3hw}{h-w}B_h -6t B_t
+\left(9w - \frac{3hw}{h-w}-\frac{3w^2}{w-z} \right)B_w
  \nonumber \\ &
  +\frac{1}{s_w^2}\left(-6w + 3z \right)B_z -\frac{2}{h}T^{(4,1)} \, .
 \end{align}
The SMEFT contribution to  $\delta v_\mu$ for arbitrary flavour structure in unitary gauge
within the FJ tadpole scheme was given in \cite{Biekotter:2023xle}, and in our notation reads
 \begin{align}
\label{eq:dVmu_SMEFT}
16\pi^2 & \delta v_\mu^{(6,1)} =
 16\pi^2   \delta v_\mu^{(4,1)}\left( -2\delta v_\mu^{(6,0)}+\frac{C_{HD} }{2}\right)   \nonumber \\ &
+ h\left(-C_{H\Box}+\frac{C_{HD}}{2}\right) 
 +5z  C_{\substack{ll \\ 1221}}
\nonumber \\ &
+w\left(-C_{H\Box}-\frac{3 C_{HD}}{2}-12\frac{s_w }{c_w}C_{HWB}
+10 C_{\substack{Hl \\ 11}}^{(3)} + 10 C_{\substack{Hl \\ 22}}^{(3)} 
+ 10   \left( C_{\substack{ll \\ 1122}} - C_{\substack{ll \\ 1221}} \right)  
\right)
\nonumber \\ &
+3t\left(-\frac{ C_{HD}}{2}
+ C_{\substack{Hl \\ 11}}^{(3)} +  C_{\substack{Hl \\ 22}}^{(3)}  
+2  C_{\substack{Hq\\ 33}}^{(3)} 
- C^{(3)}_{\substack{lq \\ 1133}}-  C^{(3)}_{\substack{lq \\ 2233}} \right)
\nonumber \\
&+6 w \frac{h B_h- w B_w}{h-w}\left( C_{H\Box}-\frac{C_{HD}}{2}\right) 
\nonumber \\ &
+6 w B_w \left( C_{\substack{Hl \\ 11}}^{(1)}  +C_{\substack{Hl \\ 22}}^{(1)} 
+C_{\substack{Hl \\ 11}}^{(3)}  +C_{\substack{Hl \\ 22}}^{(3)} +  
2 C_{\substack{ll \\ 1122}}  \right) 
\nonumber \\ &
+6 w B_z \left( -C_{HD}-C_{\substack{Hl \\ 11}}^{(1)}  -C_{\substack{Hl \\ 22}}^{(1)} 
+C_{\substack{Hl \\ 11}}^{(3)}  +C_{\substack{Hl \\ 22}}^{(3)} +  
\left(-2+\frac{z}{w}\right) C_{\substack{ll \\ 1221}}  \right) 
\nonumber \\ &
+t B_t\left(3 C_{HD}- 6 C_{\substack{Hl \\ 11}}^{(3)}  -6C_{\substack{Hl \\ 22}}^{(3)}  -12C_{\substack{Hq \\ 33}}^{(3)} 
+6 C^{(3)}_{\substack{lq \\ 1133}}+ 6 C^{(3)}_{\substack{lq \\ 2233}}\right)
\nonumber \\
&-\frac{2}{h}\left[T^{(4,1)}\left(C_{H,{\rm kin}} -\frac{1}{2}C_{HD}\right)+ T^{(6,1)}\right] \, .
\end{align}

In the $v_\alpha$ scheme, the one-loop SM result for vev renormalisation takes the form
\begin{align}
\delta v_\alpha^{(4,1)} & = 2\left(\delta M_W^{(4,1)}+ \delta s_w^{(4,1)} -\delta e^{(4,1)}\right) \, , 
\end{align}
where all pieces have been given explicitly above with the exception of 
\begin{align}
 16\pi^2 \delta e^{(4,1)} & =  
4 w s_w^2\left(\frac{283}{54}+\frac{8}{9}B_t-\frac{7}{2}B_w +\frac{40}{9}B_z\right) \,.
 \end{align}
The dimension-six piece can be written as
\begin{align}
\delta v_\alpha^{(6,1)} & = 2\left(\delta M_W^{(6,1)}+ \delta s_w^{(6,1)} -\delta e^{(6,1)}\right) 
- 2\delta v_\alpha^{(6,0)}\left(\delta M_W^{(4,1)}+ \delta s_w^{(4,1)} \right) 
\nonumber \\
& +\frac{2}{c_w s_w}\left(C_{HWB}+\frac{c_w}{2s_w}C_{HD}\right)\delta s_w^{(4,1)} \, ,
\end{align}
where in this case the new piece is
 \begin{align}
16\pi^2  \delta e^{(6,1)}  &= - 8\frac{M_W s_w}{v_\alpha} 
\left(2 \sqrt{2} m_t v_\alpha
\left(-1+B_t\right)\left(c_wC_{tB}  + s_wC_{tW}    \right)  
+9s_w  w\left(-1+B_w\right)C_W \right)
\nonumber \\ &
+ 
hB_h c_{h\gamma\gamma} +  wc_w s_w\left(16-12B_w\right)C_{HWB} - 
\frac{2}{h}c_{h\gamma\gamma}T^{(4,1)} \, ,
\end{align}
with
\begin{align}
c_{h\gamma\gamma}&= c_w^2 C_{HB}+ s_w^2 C_{HW} - c_w s_w C_{HWB}\,.
\end{align}

\newpage
\begin{table}
\begin{center}
\small
\begin{minipage}[t]{4.4cm}
\renewcommand{\arraystretch}{1.5}
\begin{tabular}[t]{c|c}
\multicolumn{2}{c}{$1:X^3$} \\
\hline
$Q_G$                & $f^{ABC} G_\mu^{A\nu} G_\nu^{B\rho} G_\rho^{C\mu} $ \\
$Q_{\widetilde G}$          & $f^{ABC} \widetilde G_\mu^{A\nu} G_\nu^{B\rho} G_\rho^{C\mu} $ \\
$Q_W$                & $\epsilon^{IJK} W_\mu^{I\nu} W_\nu^{J\rho} W_\rho^{K\mu}$ \\ 
$Q_{\widetilde W}$          & $\epsilon^{IJK} \widetilde W_\mu^{I\nu} W_\nu^{J\rho} W_\rho^{K\mu}$ \\
\end{tabular}
\end{minipage}
%
\begin{minipage}[t]{2.5cm}
\renewcommand{\arraystretch}{1.5}
\begin{tabular}[t]{c|c}
\multicolumn{2}{c}{$2:H^6$} \\
\hline
$Q_H$       & $(H^\dag H)^3$ 
\end{tabular}
\end{minipage}
\begin{minipage}[t]{4.9cm}
\renewcommand{\arraystretch}{1.5}
\begin{tabular}[t]{c|c}
\multicolumn{2}{c}{$3:H^4 D^2$} \\
\hline
$Q_{H\Box}$ & $(H^\dag H)\Box(H^\dag H)$ \\
$Q_{H D}$   & $\ \left(H^\dag D_\mu H\right)^* \left(H^\dag D_\mu H\right)$ 
\end{tabular}
\end{minipage}
%
\begin{minipage}[t]{2.5cm}
\renewcommand{\arraystretch}{1.5}
\begin{tabular}[t]{c|c}
\multicolumn{2}{c}{$5: \psi^2H^3 + \hbox{h.c.}$} \\
\hline
$Q_{eH}$           & $(H^\dag H)(\bar l_p e_r H)$ \\
$Q_{uH}$          & $(H^\dag H)(\bar q_p u_r \widetilde H )$ \\
$Q_{dH}$           & $(H^\dag H)(\bar q_p d_r H)$\\
\end{tabular}
\end{minipage}

\begin{minipage}[t]{4.7cm}
\renewcommand{\arraystretch}{1.5}
\begin{tabular}[t]{c|c}
\multicolumn{2}{c}{$4:X^2H^2$} \\
\hline
$Q_{H G}$     & $H^\dag H\, G^A_{\mu\nu} G^{A\mu\nu}$ \\
$Q_{H\widetilde G}$         & $H^\dag H\, \widetilde G^A_{\mu\nu} G^{A\mu\nu}$ \\
$Q_{H W}$     & $H^\dag H\, W^I_{\mu\nu} W^{I\mu\nu}$ \\
$Q_{H\widetilde W}$         & $H^\dag H\, \widetilde W^I_{\mu\nu} W^{I\mu\nu}$ \\
$Q_{H B}$     & $ H^\dag H\, B_{\mu\nu} B^{\mu\nu}$ \\
$Q_{H\widetilde B}$         & $H^\dag H\, \widetilde B_{\mu\nu} B^{\mu\nu}$ \\
$Q_{H WB}$     & $ H^\dag \sigma^I H\, W^I_{\mu\nu} B^{\mu\nu}$ \\
$Q_{H\widetilde W B}$         & $H^\dag \sigma^I H\, \widetilde W^I_{\mu\nu} B^{\mu\nu}$ 
\end{tabular}
\end{minipage}
%
\begin{minipage}[t]{5.2cm}
\renewcommand{\arraystretch}{1.5}
\begin{tabular}[t]{c|c}
\multicolumn{2}{c}{$6:\psi^2 XH+\hbox{h.c.}$} \\
\hline
$Q_{eW}$      & $(\bar l_p \sigma^{\mu\nu} e_r) \sigma^I H W_{\mu\nu}^I$ \\
$Q_{eB}$        & $(\bar l_p \sigma^{\mu\nu} e_r) H B_{\mu\nu}$ \\
$Q_{uG}$        & $(\bar q_p \sigma^{\mu\nu} T^A u_r) \widetilde H \, G_{\mu\nu}^A$ \\
$Q_{uW}$        & $(\bar q_p \sigma^{\mu\nu} u_r) \sigma^I \widetilde H \, W_{\mu\nu}^I$ \\
$Q_{uB}$        & $(\bar q_p \sigma^{\mu\nu} u_r) \widetilde H \, B_{\mu\nu}$ \\
$Q_{dG}$        & $(\bar q_p \sigma^{\mu\nu} T^A d_r) H\, G_{\mu\nu}^A$ \\
$Q_{dW}$         & $(\bar q_p \sigma^{\mu\nu} d_r) \sigma^I H\, W_{\mu\nu}^I$ \\
$Q_{dB}$        & $(\bar q_p \sigma^{\mu\nu} d_r) H\, B_{\mu\nu}$ 
\end{tabular}
\end{minipage}
%
\begin{minipage}[t]{5cm}
\renewcommand{\arraystretch}{1.5}
\begin{tabular}[t]{c|c}
\multicolumn{2}{c}{$7:\psi^2H^2 D$} \\
\hline
$Q_{H l}^{(1)}$      & $(H^\dag i\overleftrightarrow{D}_\mu H)(\bar l_p \gamma^\mu l_r)$\\
$Q_{H l}^{(3)}$      & $(H^\dag i\overleftrightarrow{D}^I_\mu H)(\bar l_p \sigma^I \gamma^\mu l_r)$\\
$Q_{H e}$            & $(H^\dag i\overleftrightarrow{D}_\mu H)(\bar e_p \gamma^\mu e_r)$\\
$Q_{H q}^{(1)}$      & $(H^\dag i\overleftrightarrow{D}_\mu H)(\bar q_p \gamma^\mu q_r)$\\
$Q_{H q}^{(3)}$      & $(H^\dag i\overleftrightarrow{D}^I_\mu H)(\bar q_p \sigma^I \gamma^\mu q_r)$\\
$Q_{H u}$            & $(H^\dag i\overleftrightarrow{D}_\mu H)(\bar u_p \gamma^\mu u_r)$\\
$Q_{H d}$            & $(H^\dag i\overleftrightarrow{D}_\mu H)(\bar d_p \gamma^\mu d_r)$\\
$Q_{H u d}$ + h.c.   & $i(\widetilde H ^\dag D_\mu H)(\bar u_p \gamma^\mu d_r)$\\
\end{tabular}
\end{minipage}

\vspace{0.25cm}

\begin{minipage}[t]{4.75cm}
\renewcommand{\arraystretch}{1.5}
\begin{tabular}[t]{c|c}
\multicolumn{2}{c}{$8:(\bar LL)(\bar LL)$} \\
\hline
$Q_{ll}$        & $(\bar l_p \gamma_\mu l_r)(\bar l_s \gamma^\mu l_t)$ \\
$Q_{qq}^{(1)}$  & $(\bar q_p \gamma_\mu q_r)(\bar q_s \gamma^\mu q_t)$ \\
$Q_{qq}^{(3)}$  & $(\bar q_p \gamma_\mu \sigma^I q_r)(\bar q_s \gamma^\mu \sigma^I q_t)$ \\
$Q_{lq}^{(1)}$                & $(\bar l_p \gamma_\mu l_r)(\bar q_s \gamma^\mu q_t)$ \\
$Q_{lq}^{(3)}$                & $(\bar l_p \gamma_\mu \sigma^I l_r)(\bar q_s \gamma^\mu \sigma^I q_t)$ 
\end{tabular}
\end{minipage}
\begin{minipage}[t]{5.25cm}
\renewcommand{\arraystretch}{1.5}
\begin{tabular}[t]{c|c}
\multicolumn{2}{c}{$8:(\bar RR)(\bar RR)$} \\
\hline
$Q_{ee}$               & $(\bar e_p \gamma_\mu e_r)(\bar e_s \gamma^\mu e_t)$ \\
$Q_{uu}$        & $(\bar u_p \gamma_\mu u_r)(\bar u_s \gamma^\mu u_t)$ \\
$Q_{dd}$        & $(\bar d_p \gamma_\mu d_r)(\bar d_s \gamma^\mu d_t)$ \\
$Q_{eu}$                      & $(\bar e_p \gamma_\mu e_r)(\bar u_s \gamma^\mu u_t)$ \\
$Q_{ed}$                      & $(\bar e_p \gamma_\mu e_r)(\bar d_s\gamma^\mu d_t)$ \\
$Q_{ud}^{(1)}$                & $(\bar u_p \gamma_\mu u_r)(\bar d_s \gamma^\mu d_t)$ \\
$Q_{ud}^{(8)}$                & $(\bar u_p \gamma_\mu T^A u_r)(\bar d_s \gamma^\mu T^A d_t)$ \\
\end{tabular}
\end{minipage}
\begin{minipage}[t]{4.75cm}
\renewcommand{\arraystretch}{1.5}
\begin{tabular}[t]{c|c}
\multicolumn{2}{c}{$8:(\bar LL)(\bar RR)$} \\
\hline
$Q_{le}$               & $(\bar l_p \gamma_\mu l_r)(\bar e_s \gamma^\mu e_t)$ \\
$Q_{lu}$               & $(\bar l_p \gamma_\mu l_r)(\bar u_s \gamma^\mu u_t)$ \\
$Q_{ld}$               & $(\bar l_p \gamma_\mu l_r)(\bar d_s \gamma^\mu d_t)$ \\
$Q_{qe}$               & $(\bar q_p \gamma_\mu q_r)(\bar e_s \gamma^\mu e_t)$ \\
$Q_{qu}^{(1)}$         & $(\bar q_p \gamma_\mu q_r)(\bar u_s \gamma^\mu u_t)$ \\ 
$Q_{qu}^{(8)}$         & $(\bar q_p \gamma_\mu T^A q_r)(\bar u_s \gamma^\mu T^A u_t)$ \\ 
$Q_{qd}^{(1)}$ & $(\bar q_p \gamma_\mu q_r)(\bar d_s \gamma^\mu d_t)$ \\
$Q_{qd}^{(8)}$ & $(\bar q_p \gamma_\mu T^A q_r)(\bar d_s \gamma^\mu T^A d_t)$\\
\end{tabular}
\end{minipage}

\vspace{0.25cm}

\begin{minipage}[t]{3.75cm}
\renewcommand{\arraystretch}{1.5}
\begin{tabular}[t]{c|c}
\multicolumn{2}{c}{$8:(\bar LR)(\bar RL)+\hbox{h.c.}$} \\
\hline
$Q_{ledq}$ & $(\bar l_p^j e_r)(\bar d_s q_{tj})$ 
\end{tabular}
\end{minipage}
\begin{minipage}[t]{5.5cm}
\renewcommand{\arraystretch}{1.5}
\begin{tabular}[t]{c|c}
\multicolumn{2}{c}{$8:(\bar LR)(\bar L R)+\hbox{h.c.}$} \\
\hline
$Q_{quqd}^{(1)}$ & $(\bar q_p^j u_r) \epsilon_{jk} (\bar q_s^k d_t)$ \\
$Q_{quqd}^{(8)}$ & $(\bar q_p^j T^A u_r) \epsilon_{jk} (\bar q_s^k T^A d_t)$ \\
$Q_{lequ}^{(1)}$ & $(\bar l_p^j e_r) \epsilon_{jk} (\bar q_s^k u_t)$ \\
$Q_{lequ}^{(3)}$ & $(\bar l_p^j \sigma_{\mu\nu} e_r) \epsilon_{jk} (\bar q_s^k \sigma^{\mu\nu} u_t)$
\end{tabular}
\end{minipage}
\end{center}
\caption{\label{op59}
The 59 independent baryon number conserving dimension-six operators built from Standard Model fields, in 
the notation of \cite{Jenkins:2013zja}.  The subscripts $p,r,s,t$ are flavour indices, and $\sigma^I$ are Pauli
matrices.}
\end{table}
\FloatBarrier

\bibliography{literature}
\bibliographystyle{JHEP.bst}

\end{document}